# THE EXTENSION OF THE TRANSITION TEMPERATURE PLASMA INTO THE LOWER GALACTIC HALO

## Blair D. Savage and Bart P. Wakker

Department of Astronomy, University of Wisconsin, 475 N. Charter St, Madison, WI 53706; savage@astro.wisc.edu, wakker@astro.wisc.edu.

## ABSTRACT

Column densities for H I, Al III, Si IV, C IV, and O VI toward 109 stars and 30 extragalactic objects have been assembled to study the extensions of these species away from the Galactic plane into the Galactic halo. H I and Al III mostly trace the warm neutral and warm ionized medium, respectively, while Si IV, C IV and O VI trace a combination of warm photoionized and collisionally ionized plasmas. The much larger object sample compared to previous studies allows us to consider and correct for the effects of the sample bias that has affected earlier but smaller surveys of the gas distributions. We find Si IV and C IV have similar exponential scale heights of 3.2(+1.0, -0.6) and 3.6(+1.0, -0.8) kpc. The scale height of O VI is marginally smaller with h = 2.6±0.6 kpc. The transition temperature gas is ~3 times more extended than the warm ionized medium traced by Al III with h = 0.90(+0.62, -0.33) kpc and ~12 times more extended than the warm neutral medium traced by H I with h = 0.24 ±0.06 kpc. There is a factor of 2 decrease in the dispersion of the log of the column density ratios for transition temperature gas for lines of sight in the Galactic disk compared to extragalactic lines of sight through the entire halo. The observations are compared to the predictions of the various models for the production of the transition temperature gas in the halo.

*Subject Headings:* Galaxy:halo-ISM:clouds-ISM:structure-Ultraviolet:ISM

# 1. INTRODUCTION

The Galaxy has a gaseous atmosphere consisting of different phases of the interstellar medium (ISM) including the cold neutral medium (CNM), the warm neutral medium (WNM), the warm ionized medium (WIM), and the hot ionized medium (HIM). The different phases exhibit different extensions away from the Galactic mid-plane into the halo. A study of the extensions provides information about the different phases, their different origins, and the way in which the phases interact with each other.

Absorption line spectroscopy toward objects with known distances provides an important method for measuring the extension of gas away from the Galactic plane. One can study the absorption to stars situated at different distances, d, and determine column densities, $N(X)$, for a wide variety of interstellar species, X. With a large enough sample of measured values of $N(X)$ to stars, it is possible to determine the distribution of the gas away from the Galactic plane by comparing observations of $N(X)$ and its dependence on d, l and b, to simple models of the expected distribution. It is most common to assume a plane parallel model for the distribution with a simple exponential decline away from the plane with $n(z) = n_0 \exp(-|z|/h)$ where $n_o$ is the mid-plane density, h is the exponential scale height, and $z = d \sin b$ is the distance away from the plane. In this case, the model column density integrated along the z axis, $N = \int n(z)dz = n_0 h[1-\exp(-|z|/h)]$ can be compared to the projection of the observed column densities onto the z axis, $N(X)\sin|b|$. The simple model predicts $N(X)\sin|b| = n_0 h[1-\exp(-|z|/h)]$. Plots of $\log[N(X)\sin|b|]$ versus $\log|z|$ are expected to reveal an increase $\log[N(X)\sin|b|]$ with $\log|z|$ until $\log|z| \sim \log h$. For larger $\log|z|$, $\log[N(X)\sin|b|]$ should begin to approach

log($n_0$h). This simple exponential model assumes the gas has a smooth distribution. The irregularity in the actual gas distribution will produce a scatter of the data points about the simple exponential distribution model. The scatter produced by the gas irregularity will often be much larger than the scatter produced by the observational errors.

A problem with the absorption line method described above for determining the distribution of gas is that the sample of measurements must be relatively large in order to begin to deal adequately with the effects of the irregularities in the gas distribution and the effects of sample bias on the results. With only a few exceptions such as Na I, Ca II, and Ti II, the absorption observations must generally be done in the ultraviolet region of the spectrum in order to record the relevant absorption lines of species that sample the different phases of the ISM. While results for H I (Diplas & Savage 1994) , Ti II (Lipman & Pettini 1995), Al III (Savage, Edgar & Diplas 1990), and Si IV, C IV, N V (Savage, Sembach & Lu 1997), have become available in recent years, the sample sizes for these earlier measurements have been relatively small, except for H I. With the successful operation of the FUSE satellite over the period 1999-2007, it has been possible to produce an absorption line data set tracing gas in the Galactic disk and halo for O VI toward a large number of Galactic (Bowen et al. 2008; Zsargó et al. 2003) and extragalactic objects (Savage et al. 2003). This provides the opportunity to intercompare the extension of O VI away from the Galactic plane to the extension of other important interstellar species such as H I, Al III, Si IV C IV, and N V.

The species Si IV, C IV, N V and O VI are particularly important for interstellar studies because they can be used to trace gas that cools relatively rapidly in the temperature range from $(\sim 1\text{-}6) \times 10^5$ K. This "transition temperature" gas differs from the more stable warm and hot gas phases with T ~ 5000-10,000 K and T > $10^6$ K , respectively[1]. The transition temperature gas may arise in the conductive or turbulent interface regions between warm and hot gas or in cooling gas in a Galactic fountain or in old supernovae bubbles. The best tracer of transition temperature gas is O VI because it is unlikely that much O VI in the Milky Way disk and halo is created by photoionization in warm gas except in the vicinity of very hot white dwarfs with T > 50,000 K. This is because photons with E >113 eV are required to convert O V into O VI. In contrast, Si IV and C IV be created with photons having E > 33 and >54 eV, respectively. Therefore, much of the Si IV and some of the C IV found in the ISM of the Milky Way could be produced by photoionization in warm gas. The ionizing radiation could be from hot Galactic stars, the intergalactic background radiation, or the radiation produced in warm-hot gas interfaces. N V, which requires E >77 eV for its production, also is a good diagnostic of transition temperature collisionally ionized gas. However, the lower abundance of N compared to O makes N V more difficult to detect than O VI, particularly in complex stellar spectra.

---

[1] Unfortunately the ISM and IGM literature has become very confusing in the use of the terms 'warm' and 'hot' when describing different thermal phases of the gas in space. ISM astronomers have traditionally used warm to describe gas with temperatures from ~ $(0.3-3)x10^4$ K and hot to describe gas with T > $10^6$ K. However, this ISM definition of hot is not uniformly applied and many ISM papers in the literature refer to hot gas with T ~ $10^5$ to $10^6$ K traced by species such as O VI, N V and C IV. To make this thermal naming system even more confusing, it is now common when discussing the IGM to refer to the cool, warm and hot IGM with T's ranging from $(0.5-3)x10^4$ K, $(1-10)x10^5$ K , and >$10^6$ K, respectively. Thus the warm-hot IGM or WHIM concerns warm gas with T = $(1-10)x10^5$ K and hot gas with >$10^6$ K. Unfortunately, many IGM researchers have not adopted this revised definition for warm and continue to use warm to refer to the photoionized IGM where T ~ $(1-3)x10^4$ K. It is common to read IGM papers where the use of the term warm changes back and forth from the ISM to the new IGM definition several times. A way out of this confusing set of definitions is to retain the traditional ISM use of the terms warm and hot and introduce the term "transition temperature" to refer to gas with temperatures in between warm and hot with T ~ $(0.5-10)x10^5$ K. The transition temperature gas is thermally unstable, while the warm gas with T ~$(0.3-3)x10^4$ K is often stable and the hot gas with T > $10^6$ K is relatively stable because it cools very slowly. Gas caught in the unstable intermediate temperature range is generally in transition from the hot phase to the warm phase (or from the warm to the hot phase) depending on the nature of the cooling and heating processes. The term transition helps to convey important physical information about the gas in the unstable thermal regime.

Table 1 summarizes published values of the distributions of a number of interstellar species including H I, Ti II, Al III, Si IV, C IV, N V, O VI and electrons derived from the $N(X)\sin|b|$ versus $|z|$ method. The various columns in the Table give the species, the number of Galactic and extragalactic sight lines used to derive h, $\sigma_p$(dex), a logarithmic measure of the patchiness or irregularity in the gas, $n_o \pm \sigma$, $\log(n_o h) \pm \sigma$, and $h(kpc) \pm \sigma$. The sources of the measurements and comments are found in the footnotes to the table. The relatively small sample size of the measurements for Al III, Si IV, and C IV from Savage et al. (1997) makes it unreliable to directly compare those scale height estimates with the much larger sample results for O VI, particularly given the finding that the distribution of O VI in the north Galactic polar region with b > 45° exhibits a 0.25 dex excess compared to the rest of the sky (Savage et al. 2003). Including or excluding this region when measuring h(O VI) affects the derived scale height considerably and presumably the same problem could occur for Si IV, C IV and N V and other species. When trying to interrelate the Galactic distribution of different species, it is important to obtain measurements along the same lines of sight for the different species to minimize the possible impact of sample bias on the result.

All the results listed in Table 1 are based on ultraviolet (H I, Al III, Si IV, C IV and O VI) or optical (Ti II) absorption line data with the exception of the results for the electron scale height which are derived from pulsar dispersion measures. Gaensler et al. (2008) have recently proposed that the electron galactic scale height is 1.89(+0.12, -0.25) kpc which is ~2 times larger than values previously reported. This result is derived for 15 lines of sight to pulsars with |b| > 40°. However, the fit process does not allow for the irregular (patchy) nature of the WIM and the quoted value of the reduced chi square implies the fit is poor. In fact the scale height derived by Gaensler et al. is based on an invalid fit to the pulsar observations as discussed in the appendix. A revised fit for the 15 pulsars with |b| > 40° yields a smaller scale height, h = 1.41(+0.26, -0.21) kpc.

In anticipation of the major FUSE surveys of O VI, Savage, Meade & Sembach (2001), determined column densities for Al III, Si IV and C IV toward 164 early-type stars in the Galactic disk and low halo from spectra obtained with the International Ultraviolet Explorer (IUE) satellite at a resolution of 25 km s$^{-1}$. OVI toward many of those stars were subsequently observed by FUSE at a resolution of 20 km s$^{-1}$ by Bowen et al. (2008). In this paper we bring the two data sets together for the first time.

We have supplemented the IUE and FUSE disk/halo star effort with new measurements of Si IV and C IV absorption from the Space Telescope Imaging Spectrograph (STIS) archive of 25 extragalactic objects for which good quality O VI data exist from FUSE. These new observations are from Wakker et al. (in preparation). Extragalactic observations are essential to derive the values of h for species where the largest values of |z| in the stellar sample are comparable to or smaller than h. The observations of the high ions and Al III along with measures of N(H I) toward most of the stars by Diplas & Savage (1994) and toward the AGNs from Wakker et al. (2003) makes it possible for the first time with a large sample of measurements to directly and reliably compare the relative distributions of H I, Al III, Si IV, C IV and O VI away from the plane of the Milky Way. H I mostly traces the WNM with some contributions from diffuse clouds associated with the CNM. Al III traces the WIM. Si IV and C IV trace a combination of photoionionzed gas and transition temperature gas. O VI mostly traces collisionally ionized transition temperature gas.

## 2. THE MEASUREMENTS

We have collected in Table 2 column densities for H I, Al III, Si IV, C IV and O VI for a large number of Galactic and extragalactic objects. The sources of these measurements for 109 stars in the Milky Way disk and halo are discussed in §2.1. The sources for these measurements for 25 AGNs and several stars in the Magellanic Clouds are discussed in §2.2. The basic requirement for an object appearing in Table 2 is that good measures of N(O VI) are available from the FUSE or the Copernicus satellites along with reliable measures or limits for Al III, Si IV, or C IV from either IUE or HST. Along extragalactic lines of sight very few Al III measurements exist. Therefore, it was necessary to include several directions with no measurements of O VI to study the behavior of Al III at large |z|. The large size of our sample allows us to more reliably determine the extensions of the different species away from the galactic plane and to determine those extensions for the different ions for the same lines of sight. This reduces the effect of sample bias when inter-comparing the different scale height results.

### 2.1 Sight Lines in the Disk and Halo to Galactic Stars

For the first 109 objects of Table 2 we have collected values of N(O VI) and other column densities measured along lines of sight in the Galactic disk and halo toward Galactic stars. We started with the O VI survey of Bowen et al. (2008) which includes measurements from FUSE of stars in the disk along with FUSE halo star measurements from Zsargó et al. (2003) and an updated version of the Copernicus O VI survey of Jenkins (1978). Our goal was to find objects with good O VI measurements for which measurements from other surveys exist for Al III, Si IV and C IV. Our primary source of the column densities for these three ions is the IUE high ion disk and low halo survey of Savage, Meade & Sembach (2003) that has a strong object overlap with the FUSE Bowen et al. (2008) survey. In assembling the data set of Table 2 we rejected objects for which the O VI logarithmic errors exceed 0.4 dex. We also required that reasonable quality measurements (errors <0.4 dex) exist for one or more of the ions Al III, Si IV and C IV when O VI is detected.

The various entries in the numbered columns of Table 2 include: (1) the star HD number or alternate name, (2) the MK spectral type from Bowen et al. (2008) and references therein, (3) Galactic longitude, (4) Galactic latitude, (5) stellar distance, d (kpc), from Bowen et al., (6) the estimated logarithmic error (dex) in the distance from Bowen et al., (7) prominent H II regions associated with the star as identified via the footnotes, (8)[R], a measure of the 0.25-1.0 keV brightness of the X-ray sky in the direction of each object from Bowen et al. with R = 0, 1, 2, and 3 implying an increasing brightness of the X-ray sky in the immediate vicinity of the star or AGN. (9) to (13) logarithmic column densities and errors for H I, Al III, Si IV, C IV and O VI. (14) the source of the Al III, Si IV and C IV column densities, (15) the source of the O VI column densities. The H I column densities for the stars are from the Lyman α observations of Diplas & Savage (1994). For the AGNs, the H I column densities are from the 21 cm results compiled by Wakker et al. (2003).

Bowen et al. (2008) derived O VI column densities using the profile fit (PF) method and the apparent optical depth method (AOD). Savage et al. (2001) derived Al III, Si IV and C IV column densities using the curve of growth (COG) method and the AOD method. The different methods yielded similar results in each of the two publications. However, to make the combined data set internally self-consistent, we have in most cases adopted the AOD results for Table 2 from each of the two publications. In the case of O VI the AOD column density is based on the measurement of the O VI λ1031.93 line. For Al III, Si IV, and C IV AOD column densities for each line of the respective doublets were combined using the method of Savage & Sembach (2001) to correct for the effects of unresolved saturation. The 1σ errors in the column densities are based on those derived in the various sources of the measurements. In the case of O VI, the errors allow for photon statistics and the continuum placement uncertainty but do not include a saturation correction uncertainty. In the case of Al III, Si IV and C IV, in addition to the photon statistical error and continuum placement error, the uncertainty associated with the saturation correction is included (see Savage et al. 2001 for detailed explanations). The errors originally derived were often asymmetric. However, for simplicity we report the average logarithmic error in Table 2. This is not a problem since we will see in §4 that the observational errors in the column densities are small compared to the typical sight-line to sight-line variation found in the highly ionized gas that arises from the irregular distribution of these species in the ISM. Also, when deriving gas distributions, the column density error is generally smaller than the error introduced by the uncertainty in the distance to the star. Columns 14 and 15 in Table 2 list the sources of the various column densities. In a small number of cases, instead of adopting column densities from the IUE satellite we have instead drawn on more accurate column densities found in published HST observations obtained with either the Goddard High Resolution Spectrograph (GHRS) or the Space Telescope Imaging Spectrograph (STIS).

The HST STIS data archive contains unpublished observations of ~30 of the stars listed in Table 2 at either 7 km s$^{-1}$ or 3.3 km s$^{-1}$ resolution. Although these observations should be superior to those obtained by the IUE satellite, we chose not to undertake the major data reduction task required to determine column densities from these data. Our project requires the much larger number of column density measurements provided by the IUE 164 star survey of Savage et al. (2001). In addition we note that the higher resolution HST STIS observations of the highly ionized species are being analyzed in a separate program being pursued by W. F. Zech , J. C Howk,  and N. Lehner at Notre Dame University.

Table 2 includes only a few O VI measurements obtained by the Copernicus satellite survey of Jenkins (1978) as updated by Bowen et al. (2008, see their Table 12). Many of those observations are quite uncertain with errors exceeding 0.4 dex. In addition, for most of the closer-by stars in the Copernicus satellite survey, the O VI column densities are small and the other high ions are usually not detectable (see Edgar & Savage 1992).

In a number of cases we list column density limits in Table 2. If a species is not detected we list the 3σ detection limit from Savage et al. (2001). We have not listed in Table 2 the relatively close-by lines of sight to stars observed by FUSE for which interstellar O VI is not  detected. In essentially all of these cases Al III, Si IV and C IV are also not detected.

Lower limits to column densities are listed in Table 2 for cases where line saturation is a problem. The limit listed is the larger value of the apparent column density for the two components of the doublet measured in the IUE high ion survey. In most cases this limit is obtained from the weaker line of the doublet.

The stellar absorption observations of Al III, Si IV, C IV and O VI generally are within the velocity range from -100 to 100 km s$^{-1}$. Although, multiple absorption components are occasionally recorded, we report the total column density along the line of sight integrated over each component. Absorption by intermediate velocity clouds (IVCs) is included in the column density measurements. High velocity clouds (HVCs) are not detected toward stars in the Galactic disk and low halo. Additional comments about IVCs and HVCs are found in §2.2.

The stellar distances listed in Table 2 range from 0.1 to 10 kpc with only two stars having d <0.8 kpc. The logarithmic distance errors listed in column d are from the very careful distance error assessment of Bowen et al. (2008) for all the stars in the FUSE O VI disk survey. Omitting the very distant globular cluster star vZ1128, we find the median and average distances to the Galactic stars in the sample are 2.6 and 2.8 kpc, respectively. The median and average values of |z| are 0.09 and 0.33 kpc, respectively.

## 2.2 Sight Lines through the Galactic Halo to Extragalactic Objects

We have collected measures of the high ion column densities along lines of sight to extragalactic objects at the end of Table 2. For an object to appear in this part of the Table we again require good quality O VI measurements from FUSE (errors < 0.4 dex) and the existence of measurements for other high ion column densities from IUE, GHRS or STIS.

Most of the measurements are drawn from an extension of the FUSE O VI survey of Wakker et al. (2003) and Savage et al (2003). This unpublished extension (Wakker et al. in preparation) will provide an analysis of Si IV, C IV, N V and O VI absorption at low and high velocity along all extragalactic lines of sight for which good quality data exist from FUSE for O VI and from STIS or GHRS for Si IV, C IV and N V. The entries in Table 2 report values of the total column densities of the high ions for gas in the Galactic thick disk. We assume the Galactic thick disk gas generally absorbs over a velocity range $|v_{LSR}| < 100$ km s$^{-1}$. The full details of the separation of thick disk and high velocity high ion absorption will be given in Wakker et al. (in preparation). However, the methods are similar to those employed in the original FUSE O VI survey along extragalactic sight lines (Wakker et al. 2003). The measurements refer to the total column density of absorbing gas with $|v_{LSR}| < 100$ km s$^{-1}$.

When resolved IVC components or low velocity components are present, the column density is summed over the components. The abundances in IVCs with $30 < |v_{LSR}| < 100$ km s$^{-1}$ are approximately solar (Richter et al. 2001; Wakker 2001) and the gas appears to represent the return flow of galactic fountain gas. It is therefore reasonable to include IVC absorbers as part of the Galactic thick disk. In contrast, the HVCs with $|v_{LSR}| > 100$ km s$^{-1}$ appear to trace a phenomena located beyond the galactic thick disk. The HVCs have abundances of ~0.1 to 0.3 solar (see summary in van Woerden & Wakker 2005), are found at large distances from the galactic plane (Wakker et al. 2007, 2008; Thom et al. 2008), and appear to be interacting with a highly extended (~100 kpc) hot (T~10$^6$ K) corona of the Milky Way (Sembach et al. 2003; Fox et al. 2004).  We therefore do not consider the HVCs to be associated with the Galactic thick disk and limit the column density measurements for the thick disk to $|v_{LSR}| < 100$ km s$^{-1}$.

Most of the Si IV and C IV measurements listed for extragalactic objects in Table 2 were obtained with the STIS intermediate resolution E140M echelle mode and have a resolution of 7 km s$^{-1}$.  The column densities were determined with the AOD method applied to both lines of the Si IV and  C IV doublets.   When there was no evidence for line saturation the final column densities listed were derived from an error weighted average of the individual measurements.  In several cases modest saturation corrections were applied following the procedures of Savage & Sembach (2001).  In the case of O VI, the column densities listed are updated from the original Wakker et al. (2003) survey by employing more recent data extraction and calibration procedures and when possible incorporating observations obtained after the submission of the earlier survey paper.  The total errors attached to these measurements were obtained using the methods employed in our earlier high ion survey publications.  They include contributions from photon statistics, fixed pattern noise, continuum placement uncertainty, and integration range uncertainty.  For  detailed discussions see Wakker et al. (2003)

Other sources of entries for extragalactic lines of sight in Table 2 include the Galactic portion of the absorption with  $v_{LSR} < 100$ km s$^{-1}$ toward  stars in the LMC  from Lehner & Howk (2007).

There are very few measurements of Al III $\lambda\lambda$1854.72, 1862.80 toward extragalactic objects because the STIS E140M observations do not extend beyond 1800 Å.  The small number of extragalactic values of N(Al III)  listed in Table 2 are based on IUE measurements of stars in the LMC or SMC from the literature.

## 3. OBJECT DISTRIBUTIONS AND SAMPLE BIAS

A major problem associated with the absorption method for the determination of the distribution of gas in the Galaxy is that measurements for a large number of lines of sight are required in order to average over the various sources of irregularity in the distribution of the gas.  However, even when the object sample is large various sources of observational bias can influence the result.  In this section we discuss these issues.

In Figure 1a and 1b  we display the distribution of the objects on the sky listed in Table 2.  Figure 1a shows an Aitoff view of the whole sky with directions to stars shown as  circles and directions to extragalactic objects shown as stars.

Figure 1b shows a view of the object lines of sight from above the Galactic plane. In the left panels the lines of sight to Galactic stars are shown as the solid lines and the lines of sight to the extragalactic objects are shown with dashed lines ending at a |z| distance of 8 kpc which corresponds to several times the high ion scale heights.  In the right panels the open circles display the projected positions of the survey stars.

In addition to directions through spiral arms and inter-arm regions we also need to be aware of possible excess contributions to highly ionized gas from the Local Bubble (LB) surrounding the Sun to a distance of ~100 pc,  excess photoionized Al III, Si IV and C IV in pronounced H II regions,  excess shock heated Si IV, C IV and O VI in young supernova remants,  and excess O VI produced in the wind  shocked circumstellar regions around high luminosity O and B stars.

Since most of the stars in our survey are at distances exceeding 1 kpc with median and average distances of 2.6 and 2.9 kpc, the effects of the contributions to the highly ionized gas absorption from the Local Bubble of hot gas is relatively small. The survey of O VI absorption toward white dwarfs of Savage & Lehner (2006) implies an average O VI column density of logN(O VI)= 13.05 at the ~0.1 kpc boundary of the Local bubble. The median and average values of logN(O VI) for all the stars listed in Table 2 are 14.15 and 14.11, respectively. Therefore the Local bubble contributes only ~ 9% of the total O VI column density to the average star in the survey. Including or excluding the amount of O VI in the Local Bubble has only a marginal effect on the scale heights derived for O VI in this paper. The only detection of the Si IV and C IV ions in the Local Bubble is toward α Vir (HD 116658). The values of logN for Si IV, C IV and O VI for this star listed in Table 2 imply the foreground Local Bubble contamination for Si IV and C IV will also hardly affect the derived scale heights for Si IV and C IV.

In order to investigate the effects of having the stellar or extragalactic lines of sight pass through or into pronounced H II regions we list in column 7 of Table 2 the names of the prominent H II regions associated with the survey stars. For the O stars, the associated H II region names are taken from the O star catalog of Goy (1980). For the B stars we performed our own independent search for the names of the associated H II regions. Note that these names only include the prominent H II regions found in the early Hα surveys of Sharpless (1957), Rodgers, Campbell & Whiteoak (1960) and Gum (1955). Several lines of sight pass through the Vela supernova remnant including the directions to HD74920 and HD 74711. When studying the general distribution of Al III,

Si IV, C IV and O VI we will not include the measurements for lines of sight contaminated by prominent regions of ionized gas associated with the background star.

Bowen et al. (2008) investigated the possible contributions of wind shocked regions by looking for excess 0.25 to 1 keV X-ray emission in the regions surrounding the stars in their survey.  Stars with R = 3  or 2 in Column 8 of Table 2 have relatively bright X-ray emission in their direction and may have an enhancement of O VI associated with a possible surrounding wind blown bubble.  R = 0 objects have weak surrounding X-ray emission and R = 1 objects represent an intermediate case.   Note that many of the cases of objects with R = 2 and 3 lie in directions where there are prominent H II regions associated with the star.   For the extragalactic objects we list R= 0 in all cases since extragalactic wind blown regions associated with these objects will produce redshifted absorption that is clearly distinguished from Galactic absorption.

Savage et al. (2003) found that the general distribution of Galactic thick disk  O VI observed toward 100 extragalactic objects exhibits a 0.25 dex enhancement in the north Galactic polar region with b > 45°.  Including or excluding this region of the sky when estimating the Galactic scale height of O VI  has a substantial effect on the result (see Table 1).  We need to be aware of this problem when determining the relative distributions of the highly ionized species.

The Galactic targets for the measurements reported in Table 2 are O and B stars that are relatively bright in the ultraviolet.  Since ultraviolet extinction due to interstellar dust is large, the object sample is biased toward directions with much less than normal reddening per unit distance and with relatively small values of the average line of sight H I density.  This is revealed by the average sample value in the mid-plane of $n_0$(H I) = 0.28

cm$^{-3}$ (see §4) which is 3.9 times smaller than the overall Galactic H I mid-plane average of 1.1 cm$^{-3}$ (Bohlin, Savage & Drake 1978) in the vicinity of the Sun. Our disk sample, therefore, mostly avoids the denser interstellar regions containing cool and cold gas. The directions sampled with the measurements listed in Table 2 mostly trace gas in the lower density WNM rather than the CNM.

Examples of other factors that could influence the results include the presence or absence of intermediate velocity clouds along the line of sight and the possible overlap in velocity space of high ionization absorption associated with high velocity clouds and the gas in the thick Galactic disk. Savage et al. (2003) studied the possible impact of intermediate velocity clouds on the amount of O VI in the Galactic thick disk and found no evidence for an enhancement or deficiency of O VI in directions with intermediate velocity clouds compared to other directions. Separating high ionization absorption associated with HVCs and the Galactic thick disk gas is often not easy because the HVC absorption can overlap the absorption by thick disk gas (Wakker et al. 2003 and Sembach et al. 2003). Therefore, there may be extragalactic directions where the gas absorption we have identified as thick disk gas is contaminated by the lower velocity gas associated with a HVC. An example is the direction to Complex C where O VI HVC absorption often extends from -100 km s$^{-1}$ to -200 km s$^{-1}$ (Wakker et al. 2003). The thick disk absorption toward Complex C typically has been assumed to lie between ~ -90 and +100 km s$^{-1}$ based on the appearance of the absorption in the velocity range of overlap. However, Complex C could have a lower velocity tail of absorption extending into the -90 to 0 km s$^{-1}$ velocity range. Since Complex C is situated ~ 4-11 kpc away from the Galactic plane  (Wakker et al. 2007),  it could be interacting with and slowed down  by the high z extension of thick disk gas.

It is possible that there is an excess of transition temperature gas in the Galactic center.   Therefore lines of sight passing over or near the Galactic center might have more highly ionized gas than elsewhere as a consequence of a wind driven by the central regions of the Galaxy (Keeney et al. 2006)

## 4. SCALE HEIGHTS

A major goal of this study is to evaluate similarities or differences in the distribution of H I, Al III, Si IV, C IV and O VI away from the Galactic plane.  Although we have assembled a relatively large data base from the perspective of ultraviolet absorption line standards, the sample is small given the complexity of the range of phenomena occurring in the Galaxy that might influence the distribution of the gas (see §3).  We will therefore need to be careful in deciding what observations should be included or excluded from the analysis in order to remove the possible bias introduced by various types of ISM phenomena.

In Figure 2 we display $\log[N(x)\sin|b|]$ vs $\log|z|$ for H I, Al III, Si IV, C IV, and O VI for all the observations listed in Table 2 .   Detections are denoted with the open or filled circles. Upper or lower limits are denoted with the downward or upward pointing open or filled triangles. Open symbols are for objects associated with pronounced nebular H II regions or SNRs or strong X-ray 0.25 to 1 keV X-ray emission (R = 2 or 3 in column 8 of Table 2).  The open symbols are therefore for directions where H II region photoionization may enhance the column densities of the transition temperature ions observed in this survey and where the X-ray emission may be revealing wind shocked regions around the observed stars. The extragalactic measurements plotted on the right hand side of the individual panels are ordered according to Galactic latitude with b = -90º to the left and b = 90º to the right. Extragalactic objects with b > 45º and the star vZ 1128 with b = 78.69º and z = 10 kpc are also plotted with open circles because they lie in the north Galactic polar direction where the projected , log(Nsin|b|), column density of O VI exhibits a 0.25 dex excess.

Most of the observed scatter in the panels of Figure 2 is produced by the irregularity in the ISM rather than observational errors. In each panel of Figure 2 we show the behavior of log$\{n_0 h[1-\exp(-|z|/h)]\}$ plotted against log|z| describing a smooth exponential gas distribution with the mid-plane densities listed in Table 3 and scale heights of 0.1, 0.3, 1, 3 and 10 kpc. A visual inspection of the different panels reveals the increase of implied scale heights through the sequence H I to Al III to the higher ions.

Since we are interested in the distribution of the transition temperature ions in the general ISM we should avoid particular directions if there is evidence for an excess of the transition temperature ions associated with phenomena local to the star. A simple visual inspection of the various panels of Figure 2 reveals that the open symbols do lie higher than the filled symbols in the plots of log[N(x)sin|b|] vs log|z| for Al III, Si IV, C IV, and O VI. Toward the Galactic stars the effect is strongest for Al III and Si IV. Since these two ions can be created by photoionization with photon energies greater than 18.8 and 33.5 eV, respectively, it is likely we are seeing an excess of Al III and Si IV in the photoionized gas of the H II regions associated with the survey stars.

After a consideration of the various sources of bias, we conclude the estimate of the Galactic scale heights of the transition temperature gas in the general ISM should omit lines of sight passing through the following regions:  1. Prominent H II regions.  2. Supernova remnants. 3. Objects with surrounding X-ray bright emission ( R = 2 or 3 in Table 2).  4. Extragalactic directions toward the north Galactic pole  with  b >45° along with the measurement for vZ 1128.   We will refer to the remaining sample of objects as the "restricted sample".

The visual inspection of Figure 2 reveals that the measurements for objects in the restricted sample and in the full sample exhibit a large dispersion about the prediction of a smooth simple exponential distribution for the gas.  An improved gas distribution model is one where the model is allowed to have a degree of irregularity or patchiness in the gas distribution but with an exponential decline in the number of clouds or structures sampled with distance away from the Galactic plane.  Therefore, when comparing the observations to the smooth exponential model, we include in the fitting process the patchiness parameter, $\sigma_P$,  to describe the irregular distribution of the gas in the improved model.  The patchiness parameter is adjusted to make the quality of the fit reasonable. This technique follows the method first introduced by Savage et al. (1990) and subsequently adopted in many of more recent derivations of scale heights from absorption line observations (see Table 1).

Values of the patchiness parameter, $\log(n_0 h)$, mid-plane density, scale height and their errors  derived from the observations are listed in Table 3 for the restricted sample. The best fit of the observed values of $N(X)\sin|b|$ to the exponential scale height model, $N(X)\sin|b| = n_0(X)[1- \exp(-|z|/h(X)]$, was performed through a $\chi^2$ minimization procedure

applied in  log[N(X)sin|b|]–log|z| space.  For the stellar lines of sight the logarithmic error for each observed value of log[N(X)sin|b|] is taken to be the quadrature addition of the logarithmic column density error, $\sigma_N$,  the logarithmic distance error, $\sigma_d$, and the logarithmic patchiness parameter, $\sigma_P$.   Therefore $\sigma = (\sigma_N^2 + \sigma_d^2 + \sigma_P^2)^{1/2}$.  During the minimization $\sigma_P$ is varied so that the value of the reduced $\chi^2$,   $\chi_\nu^2 = \chi^2/(N-2) = 1.0$. Here, N refers to the number of data points and 2 to the number of fitting parameters. By varying $\sigma_P$  to obtain  $\chi_\nu^2 = 1.0$, we achieve an acceptable fit for our exponential scale height model with an irregular (patchy) cloud  distribution where $\sigma_P$ is a measure of the patchiness.  If the patchyness parameter is not included in the $\chi^2$ minimization process, the resulting values of  $\chi_\nu^2$ (minimum) are  very large and the resulting  best fit values of h and log($n_0$h) are highly dependent on the most accurate column density measurements. When most of the scatter about the smooth exponential model is due to the irregularity of the ISM, it is incorrect to give too much weight in the fitting process to the most accurate measurements since the scatter in the measurements reveal the basic inadequacy of a simple smooth model.

In the fitting procedure we combine in quadrature the column density (y-axis) errors of the fit with the distance (x-axis) errors.  As discussed by Bowen et al. (2008), this is an acceptable way to include the distance errors in the minimization process for measurements made at heights |z| < h.  For the extragalactic lines of sight we do not include the distance error since for |z|>>h the error in distance does not influence the fit. Our procedure for including the distance error is not strictly valid for H I and Al III where a number of stars have |z| > h.  However, we do not believe this will have much of an impact on the derived value of the scale height because the value of $\sigma_P$  required to obtain $\chi_\nu^2 = 1.0$ is considerably larger than the average values of $\sigma_N$ and $\sigma_d$. This implies that the variation of the fit about the smooth exponential model is dominated by the irregular distribution of the gas and not by column density or distance errors.

The $\pm 1\sigma$ errors derived for the parameters listed in Table 3 were obtained by determining the $\chi^2$ (minimum) $+2.3$ contour in h-log($n_0$h) space. Here, 2.3 is the increase in $\chi^2$ corresponding to the 68% confidence or $\pm 1\sigma$ contour for a two parameter fit. For detailed discussions see Lampman, Margon & Bowyer (1976). Note that in some of the earlier applications of this method the 68% contour was incorrectly determined from $\chi^2$ (minimum) $+1.0$. This expression is only correct when the number of fit parameters is 1. The best fit results from Table 3 are shown in Figure 2 with the heavy lines. The $\pm 1\sigma$ errors in the scale height are displayed with the dashed lines.

For the restricted sample, the transition temperature high ions (O VI, C IV and Si IV) are found to have scale heights ranging from 2.6 to 3.6 kpc with errors of 0.5 to 1.2 kpc. The scale height of O VI, h = 2.6±0.5 kpc appears to be marginally smaller than the scale heights for Si IV and C IV, h = 3.2(+1.2, -0.6) kpc and 3.6(+1.0,-0.8) kpc, respectively. The scale heights for the transition temperature ions are ~3 times larger than for Al III, a tracer of the WIM, with h = 0.90(+0.62, -0.33) and ~12 times larger than for H I, mostly a tracer of the WNM with h = 0.24±0.06 kpc. Note that the comparison of the relative sizes of these scale heights is more reliable than the absolute values of the scale heights. Sample bias issues will affect the absolute values of the scale heights. However, the bias problems cancel to first order when evaluating the relative values of the scale heights because the samples are approximately the same for all the scale heights derived for the restricted sample.

We considered the effects of including measurements of upper and lower limits on the estimates of the scale heights. This was done by assuming the limit is a measurement with a relatively large column density error of $\sigma_N = 0.2$ dex. The inclusion of the limits did not affect the derived value of the fit parameters.

The scale heights for the transition temperature ions in this paper can be compared to the earlier results listed in Table 1. Because of the larger sample size and better control over sample bias issues the new scale heights are better suited for a comparison of any scale height differences from one species to the next. In the case of O VI, the value derived in this paper of 2.6±0.5 kpc is somewhat smaller than h = 3.2±0.8 kpc derived by Bowen et al. (2008) using measurements from the FUSE O VI disk survey combined with the FUSE O VI halo survey of Savage et al. (2003) for extragalactic objects in the southern Milky Way with b < -20°. Most of the difference between these two measures of h(O VI) arises from the smaller value estimated for the mid- plane density of O VI for the Bowen et al survey compared to this survey ($n_o = 1.33 \times 10^{-8}$ versus $1.64 \times 10^{-8}$ cm$^{-3}$). The Bowen et al. survey includes many more stars at d < 1 kpc than the survey of this paper. When considering disk stars at these smaller distances, Bowen et al included a correction for O VI absorption in the Local Bubble. Evidently the mid-plane density of O VI is somewhat larger when averaging over the larger distances toward the disk stars used in our survey sample.

The values of the patchiness parameter listed for the various ions in Table 3 provides important information about the degree of irregularity of the various gas phases sampled in the ISM. The five species sampled have patchiness parameters ranging from 0.172 (H I) to 0.273 (C IV) dex. Therefore the WNM, WIM and transition temperature gas sampled by these measurements are irregularly distributed with degrees of patchiness ranging from 49 to 87%. The patchiness parameter measures a complex combination of irregularities caused by intercepting absorbing structures having different size, shape, properties, and line of sight distributions. Note that when deriving the patchiness parameter through the scale height $\chi^2$ analysis method we assumed the level of patchiness is the same for lines of sight to stars in the disk and for the extragalactic lines of sight. In the next section we will see that there is more irregularity in the behavior of ion ratios along disk lines of sight than along extragalactic lines of sight.

The ~0.25 dex excess of log[N(O VI)sin|b|] toward the north Galactic Pole in directions with b > 45° found by Savage et al. (2003) is also seen toward the extragalactic objects in the Si IV and C IV ions (see panels in Fig. 2 and Table 4). In Table 4 the difference between < log[N(X)sin|b|]> for the samples with b < 45° versus b > 45° is 0.21, 0.19 and 0.23 dex for O VI, C IV and O VI, respectively. For our sample all three of these high ions have a similar average excess of ~0.21 dex for b > 45°. Evidently, the processes responsible for origin of the excess are not significantly affecting the relative behavior of the three high ions. We note that the excess high ionization absorption seen toward the 9 AGNs at b > 45° is also seen toward the globular cluster star vZ 1128 at z = 10 kpc in the direction l = 42.5° and b = 78.7°. For vZ 1128 the values of log[N(X)sin|b|] for the high ions are as large as the largest measured for any extragalactic object. This implies that the excess high ionization absorption for directions with b > 45° occurs at z < 10 kpc.

The north/south Galactic asymmetry in the distribution of interstellar gas as seen from the solar position in the Galaxy will influence scale height derivations for many species.

In the appendix we discuss how the asymmetry has likely affected the electron scale height reported by Gaensler et al. (2008). In that discussion we note that the asymmetry is also seen in FUSE measures of S III and Fe III which mostly trace gas in the WIM.

## 5. ION RATIOS IN THE TRANSITION TEMPERATURE PLASMA

The different models for the origins of the transition temperature plasma in the Galactic disk and halo make different predictions for the ionic ratios of Si IV, C IV and O VI. In Figure 3 we display the values of log[N(C IV)/N(O VI)], log[N(Si IV)/N(O VI)], and log[N(Si IV)/N(C IV)] as a function of log|z| and logd. The restricted sample is plotted with the filled circles. Stars in prominent H II regions, SNRs, and with strong associated surrounding X-ray emission, and extragalactic objects with b > 45° and vZ1128 are plotted as open circles. Lower and upper limits and data points for measurements with large error bars ($\sigma_N \geq 0.26$ dex) are not plotted to reduce the confusion. Table 4 gives medians, averages and standard deviations for the logarthimic ionic ratios for various samples of objects as defined in the footnotes to the table.

A visual inspection of the various panels in Figure 3 shows log[N(C IV)/N(O VI)], log[N(Si IV)/N(O VI)], and log[N(Si IV)/N(C IV)] are often larger for objects with suspected H II region or stellar wind contamination (open circles) versus directions where the line of sight does not pass through these contaminated environments. The behavior justifies the omission of those directions when determining the scale heights in §4. The column densities of Si IV and C IV and sometimes O VI are clearly enhanced for directions passing through prominent H II regions.

Examples of lines of sight with very large ratios of the Si IV to C IV column densities include HD 101131, HD 101205, HD 101298, HD 101413, and HD 101436 with <log[NSi IV)/N(C IV)]> = -0.02 dex which is 0.57 dex larger than for the restricted sample average. All 5 of these stars are situated in the pronounced H II region Gum 42 with the exciting cluster IC 2944. Deep Hα imaging of this region of the sky around (l, b) = (295º, -1.7º) is discussed by Georgelin et al. (2000). The lines of sight to HD 101190 and HDE 308803 also pass toward Gum 42. However, the C IV line is saturated for HD101190 and the Si IV line is saturated for HDE 308803. The limit log[N(Si IV)/N(C IV)] $\geq$ -0.17 for HDE 308803 is consistent with the results for the five stars discussed above.

Other lines of sight with large Si IV to C IV column density ratios are for HD 152623 and HD 152723 which are located in Gum 56 and HD 153426 in Gum 57a. These three stars have <log[NSi IV)/N(C IV)]> = -0.07 dex which is 0.52 dex larger than for the restricted sample average.

Stars in the Carina nebular complex including HD 93129a, HD 93205, HD 93206, HD 93222, HD 93146, HD 93250, CPD -59 2603, HDE 303308 also have enhanced ratios of Si IV to C IV. The Si IV lines of four of these stars are so strong they are saturated and therefore only permit the derivation of a lower limit to N(Si IV). The average lower limit to log[N(Si IV)/N(C IV)] for the four stars is $\geq$ -0.20. The other four stars with secure measures of N(Si IV) and N(C IV) have <log[N(Si IV)/N(C IV)]> = -0.26. The stars in the Carina region have Si IV to C IV column density ratios enhanced by an average of >0.39 and 0.33 dex compared to the restricted sample average.

Starlight photoionization of Si IV and to some extent C IV by the O stars in these H II regions is likely responsible for the excess Si IV and C IV and the large ratio of Si IV and C IV. The O stars also have strong winds that can create excess O VI and C IV in the surrounding wind shocked regions.

For the restricted sample there is a considerable dispersion in the measured values of the column density ratios particularly for N(C IV)/N(O VI) and N(Si IV)/N(O VI). The dispersion significantly decreases when moving from lines of sight to stars in the Galactic disk and low halo with log|z| and logd < 1 to the extragalactic lines of sight with log|z| and logd > 1. This is well illustrated for the values listed in Table 4 for the median, average and stdev of the ionic ratios when moving from the results for Galactic stars using the restricted sample to the results for the extragalactic restricted sample. In particular, <log[N(Si IV)/N(C IV)]> = -0.55±0.21(stdev) in the Galactic star restricted sample and –0.60±0.11(stdev) in the extragalactic restricted sample with b < 45°. The dispersion as measured by the stdev decreases from 0.21 to 0.11 dex. <logN[(C IV)/N(O VI)] > = -0.29±0.35(stdev) in the Galactic star restricted sample and –0.15±0.17 in the extragalactic restricted sample. In this case the stdev decreases from 0.35 to 0.17 dex.

The logarithmic dispersion in the ion ratios as measured by the standard deviation is approximately two times larger for logN[(C IV)/N(O VI)] and log[N(Si IV)/N(O VI)] than for log[N(Si IV)/N(C IV)] for both the Galactic restricted sample (stdev = 0.35 and 0.38 vs 0.21 dex) and for the extragalactic restricted sample (stdev = 0.17 and 0.19 vs 0.11 dex). Si IV and C IV are much better correlated with each other than C IV versus O VI or Si IV versus O VI. The relatively good correlation between Si IV and C IV has been known for many years (see Pettini & West 1982; Savage et al. 1994) suggesting the production and destruction of these two species are controlled by regulating processes in the general ISM of the disk and halo. However, it is also clear that strong increases occur in the Si IV to C IV ratio in dense H II regions where an excess of Si IV is created by starlight photoionization.

The average values of logN(Si IV)/N(C IV), log[N(C IV)/N(O VI)] and log[N(Si IV)/N(O VI)] listed in Table 4 for the restricted galactic and restricted extragalactic samples change for the sample of sight lines to stars in the Galactic thick disk to the extragalactic sample by -0.05, 0.14 and 0.09 dex, respectively. These changes are consistent with the expectations for species with the somewhat different scale heights of 2.6±0.5, 3.2(+1.2, -0.6) and 3.6 (+1.0, -0.8) kpc, for O VI, Si VI, and C IV respectively (see Table 3). Therefore, the differences in the scale heights for Si IV, C IV and O VI listed in Table 3 are consistent with the z distribution of the ionic ratios of the three species. This gives added support to the reality of the scale height differences found among the transition temperature ions.

## 6. IMPLICATIONS OF RESULTS

Si IV, C IV and O VI extend to roughly similar distances (h ~ 3 kpc) away from the Galactic plane. The high ion extension is ~3 times the warm ionized gas extension as traced by Al III. The high ion extension is ~12 times the extension of the WNM as traced by H I. Since the extensions of these species have been measured for the same lines of sight, we can be reasonably confident in the validity of the relative behavior of the extensions from one species to the next. The new scale heights from this paper are displayed in Figure 4 with the filled circles. The open circle for the uncertain N V scale height from Savage et al. (1997) is based on a small sample of extragalactic lines of sight.

For an isothermal atmosphere at temperature T in hydrostatic equilibrium, the Galactic scale height is given by h = kT/<m>g(|z|), where <m> is the average mass per particle and g(|z|) is the gravitational acceleration toward the disk. For |z| in the range of 1 to 10 kpc in the Solar neighborhood g(|z|) $\sim 10^{-8}$ cm s$^{-2}$ (Kalberla & Dedes 2008, see their §6.1 and Fig. 13). For an isothermal Galactic thick disk in hydrostatic equilibrium, a scale height of 3 kpc would require T $\sim 0.8 \times 10^6$ K for a fully ionized plasma with solar abundances with <m> = 0.73m$_H$. The required temperature is 4-8 times larger than the temperature where O VI and C IV peak in abundance in collisional ionization equilibrium. The existence of the transition temperature ions in the Galactic thick disk therefore requires non-thermal support mechanisms (turbulence, cosmic ray pressure, magnetic pressure) or the presence of a flow process (the Galactic fountain) driven by gas with T $> 0.8 \times 10^6$ K.

It is surprising that Si IV and C IV may have slightly larger scale heights than O VI since O VI is mostly associated with production in collisionally ionized gas while Si IV and C IV are associated with both photoionized and collisionally ionized gas. In a simple Galactic fountain flow where the plasma cools and recombines during the return flow to the disk, the scale height of the higher temperature ions (O VI) should exceed that for the lower temperature ions (C IV and Si IV).

A hybrid model describing different relative extensions of Si IV, C IV, and O VI into the halo was suggested by Ito & Ikeuchi (1988) who proposed a Galactic fountain flow of cooling hot ($10^6$ K) gas to create collisionally ionized Si IV, C IV and O VI in the disk and low halo in combination with photoionization by the UV extragalactic background and Galactic EUV sources to produce an enhanced production of Si IV and C IV at high |z|. Another hybrid model capable of explaining the somewhat more extended |z| distribution of Si IV and C IV compared to O VI is that of Shull & Slavin (1994) who combine the collective effects of the production of hot gas in cooling SN bubbles at low |z| with the turbulent mixing layers created by Galactic chimneys at large |z|. However, because of the large number of adjustable parameters in such hybrid models, it is difficult of devise tests to determine which models best describe the dominating processes operating in the transition temperature gas. Hydrodynamical simulations have begun to provide important insights about the effects of supernovae explosions and hot stellar winds on gas in the Galactic disk and low halo (de Avillez & Breitschwerdt 2005). However, the modeling of the behavior of the transition temperature plasma is difficult to implement because non-equilibrium processes strongly influence the ionization particularly for Si IV and C IV.

The large decrease in the dispersion of the high ion ratios from sight lines emphasizing disk gas to sight lines emphasizing halo gas is probably the result of moving from highly disturbed regions of the Galactic disk where the primary energy injection and ionization processes are occurring through supernova explosions, stellar winds, and O star photons to the more quiescent halo regions where direct localized energy injection processes are less common. Gaseous structures at large distances away from the Galactic plane may be larger and longer-lived than structures in the disk. This might also explain the decrease in dispersion. The possible existence of large and very long-lived supernova remnants in the halo as modeled by Shelton (1998, 2006) might produce halo gas with relatively uniform high ion ratios.

The much better correlation between the column densities of Si IV and C IV than between C IV and O VI for the gas in the restricted sample of the general ISM is interesting. As we move from Si IV to C IV to O VI, it becomes more and more difficult to create the ion by starlight photoionization. Therefore, we would expect starlight photoionization to strongly affect the behavior of Si IV along a given line of sight. Indeed lines of sight extending through pronounced H II regions in the disk have large Si IV to C IV ratios as expected. If the Si IV in the general ISM is also created by starlight photoionization, one might also expect to see large variations in the Si IV to C IV ratio in that medium as the ratio of the ionizing photon density to the gas density changes from place to place along a line of sight. The absence of large variations suggests that the ionization of much of the Si IV in the general ISM may not be due to starlight photoionization. A high resolution (3.5 km s$^{-1}$) and good S/N (22 to 38) study of the high ionization line absorption along the 4 kpc line of sight to HD 167756 in the inner Galaxy by Savage et al (1994) revealed the presence of Si IV and C IV absorbing components with properties well explained by their production in warm-hot gas conductive interfaces where the C IV occurs in cooling collisionally ionized interface gas and the Si IV is produced by the He$^+$ ionizing radiation emitted by the interface. In those components log[N(Si IV)/N(C IV)] = -0.54 close to the value -0.55 dex found for the general ISM of the disk in this paper. In this case, the physics of the interface processes provides the formation and destruction regulation that yields a standard value for the column density ratio.

Extending such high resolution studies to other lines of sight in the Galactic disk will be important in order to determine if the behavior of the highly ionized gas toward HD 166756 is common to that found for other lines of sight through the Galactic disk. In Figure 5 we combine the results for ~3.5 km s$^{-1}$ resolution observations of the Si IV and C IV absorption components studied in detail along 5 lines of sight including HD 116852 (Fox et al. 2003), HD 119608 (Sembach et al. 1997), HD 167756 (Savage et al. 1994), HD 177989 (Savage et al. 2001), and HD 215733 (Fitzpatrick & Spitzer 1997). A total of 20 components containing C IV are detected. For each of the 17 components also detected in Si IV we plot N(C IV) versus log[N(Si IV)/N( C IV)] in Figure 5. The 5 lines of sight which have a combined extent of 20.5 kpc mostly sample gas in the Galactic disk and low halo although in two cases the paths extend through special environments. The sight line to HD 177989 samples gas in the Scutum supershell at v ~ 45 km s$^{-1}$. The sight line to HD 119608 samples gas in the supershell over the Sco-Oph association which is associated with Galactic radio loop IV and the North Polar Spur. The three absorption components associated with these supershells are marked SSS and NPS in Figure 5. These supershell components have small values of log[N(Si IV)/N(C IV)] = -0.70, -0.92, and -0.99. In the remaining 14 components which trace less disturbed gas in the general ISM of the disk and low halo the values of log[N(Si IV)/N(C IV)] range from -1.63 to +0.15. The four larger column density components in this group have log[N(Si IV)/N(C IV)] between -0.75 and -0.63. Nine of the 10 low column density components with N(C IV) < 2x10$^{13}$ cm$^{-2}$ have log[N(Si IV)/N(C IV)] ranging from -0.62 to +0.15. The weak component with N(C IV) = 1.05x10$^{13}$ cm$^{-2}$ and log[N(Si IV)/N(C IV)] =-1.63±0.40 from Savage et al. (2001) has C IV and Si IV well aligned in velocity but the very different b values  [b(Si IV) = 4.3±2.8 and b(C IV) = 15.9±1.8 km s$^{-1}$ ] imply the two species do not co-exist in the same absorbing medium.

There is a large spread in the Si IV to C IV ratio in the weak components.  The weak components with values of log[N(Si IV)/N(C IV)] = 0±0.2 likely arise in photoionized gas as confirmed by the narrow C IV line widths (Fox et al. 2003).  In Figure 5 the four components with narrow C IV line profiles implying gas with T <5x10$^4$ K are plotted as the  filled squares.  See Fox et al. (2003) for additional discussions.

Our brief examination of the results for 5 lines of sight implies that even though Si IV and C IV seem relatively well regulated when viewed at 25 km s$^{-1}$ resolution, the situation looks more complex at 3.5 km s$^{-1}$ resolution.  While there may be a regulation process affecting the stronger components, a wide range of ionic ratios are found in the weak components.  A more extensive high resolution study of Si IV and C IV absorption would help to determine the relative frequency of occurrence of strong components with log[N(Si IV)/N(C IV)]  near -0.6 dex versus  the weaker components exhibiting a large spread in the values of log[N(Si IV)/N(C IV)].

Toward the north  Galactic polar region in directions with b > 45° there is a 0.20 dex excess in the amounts of O VI,  C IV  and Si IV.  This excess likely occurs in the thick disk of the Galaxy with z < 10 kpc since a similar excess is found along the line of sight to the star  vZ 1128 in the Globular cluster M3 at z = 10 kpc.   The excess is possibly associated with a high z extension of the local superbubbble traced by radio loops I and IV and powered by the Sco-Cen OB association.   Other possibilities include an excess produced by a low velocity tail of absorption associated with high velocity cloud Complex C  or  an excess that  traces an outflow of transition temperature plasma from the center of the Galaxy. Origins occurring beyond the Galactic thick disk appear ruled out by the vZ 1128 measurements.

## 7. SUMMARY

We have assembled absorption line column densities of H I, Al III, Si IV, C IV and O VI tracing gas in the Galactic disk and halo toward 109 Galactic stars, 25 AGNs, and 5 Magellanic Cloud stars. New column densities for C IV and Si IV are presented for the AGNs. The measurements are drawn from observations with the IUE, FUSE, and HST observatories. The observations are used to determine the relative extensions away from the Galactic plane of a tracer of the WNM (H I), the WIM (Al III) and tracers of warm photoionized and transition temperature gas (Si IV, C IV, O VI). Our sample size is much larger than for previous studies of the relative extensions of these species into the halo. The large sample size for absorption line measurements allows an evaluation of the effects of various types of galactic phenomena that can seriously bias the results for small samples. The results are:

1. The transition temperature ions traced by Si IV, C IV and O VI exhibit similar extensions away from the Galactic plane with exponential scale heights h ~3 kpc. The scale height for O VI (h = 2.6±0.5 kpc) appears to be marginally smaller than the Si IV and C IV scale heights of h = 3.2(+1.2, -0.6) and h = 3.6 (+1.0, -0.8) kpc, respectively.

2. The transition temperature plasma is ~3 times more extended than the WIM with h = 0.90(+0.62, -0.33) kpc as traced by Al III.

3. The transition temperature plasma is ~12 times more extended than the WNM with h ~ 0.24±0.06 kpc as traced by H I.

4. The marginally smaller scale height of O VI compared to Si IV and C IV is consistent with the predictions of several hybrid halo gas models that incorporate collisional ionization in cooling transition temperature gas with photoionization by the extragalactic background or the turbulent mixing of warm and hot gas in Galactic supershells.

5. All of the species studied exhibit an irregular (patchy) distribution about a simple uniform plane parallel exponential model for the extension of the gas away from the Galactic plane. The level of the irregularity as measured by the patchyness parameter ($\sigma_P$) ranges from 0.172 (H I) to 0.273(C IV) dex. The measured dispersion of the observations about the smooth exponential model is dominated by the irregular distribution of the gas and not by column density or distance errors.

6. There is a factor of ~2 decrease in the logarithmic dispersion of the ionic column density ratios for the transition temperature plasma in moving from lines of sight in the Galactic disk to lines of sight through the entire halo.

7. The dispersions for log[N(C IV)/N(O VI)] and log[N(Si IV)/N(O VI)] in the disk and halo are ~2 times that for log[N(Si IV)/N(C IV)]. Si IV and C IV are relatively tightly correlated with each other with <log[N(Si IV)/N(C IV)]> = -0.55±0.21 (stdev) and –0.60±0.11 (stdev) in the disk and halo, respectively.

8. The assembled set of observations reveals lines of sight along which the high ions have properties deviating significantly from the average results discussed above. These include lines of sight through prominent H II regions where there is a strong enhancement in the amount of Si IV compared to C IV.

9. Toward the north Galactic polar region in directions with b > 45º there is a 0.20-0.25 dex excess in the amounts of O VI, C IV and Si IV. This excess likely occurs in the thick disk of the Galaxy with z < 10 kpc since a similar excess is found along the line of sight to the star vZ 1128 in the globular cluster M3 at z = 10 kpc.

10. In the appendix we discuss a revision to the Galactic electron scale height recently proposed by Gaensler et al. (2008) based on electron dispersion measures. The revised result is strongly influenced by the north/south Galactic asymmetry found for many interstellar species. More dispersion measures are needed for pulsars with b < -45º to determine the actual size of the electron asymmetry and its effect on the inferred value of the electron scale height.

The results reported here used observations from a number of NASA observatories and instruments including the Copernicus satellite, IUE, FUSE, HST-GHRS, and HST-STIS. We appreciate the dedication and hard work required to plan, build and successfully operate these facilities that span nearly four decades of space astronomy. We thank the referee for providing a number of suggestions that helped to improve the manuscript. The Si IV and C IV measurements obtained toward the extragalactic objects were mostly obtained with STIS aboard the NASA/ESA HST which is operated by the Association of Universities for Research in Astronomy, Inc. under NASA contract NAS5-25655. The O VI results were obtained by the NASA-CNES FUSE mission operated by Johns Hopkins University. Financial support to US participants has been provided by NASA contract NAS5-32985.

References

Bohlin, R. C., Savage, B. D., & Drake, J. F. 1978, ApJ, 224, 132

Bowen, D., Jenkins, E. B., Tripp, T. M. et al. 2008, ApJS, 176, 59

Brandt, J. et al. 1999, ApJ, 117, 400

de Avillez, M. A., & Breitschwerdt, D.,  2005,  ApJ,  634, L65

Diplas, A., & Savage, B. D. 1994, ApJ,  427, 274

Edgar, R., & Savage, B. D. 1992, ApJ, 396, 124

Fitzpatrick, E. L., & Spitzer, L. 1997, ApJ,  475, 623

Fox, A.J., Savage, B. D., Sembach, K. R., Fabian, D., Richter, P., Meyer, D. M.
    Lauroesch, J., & Howk, J. C.  2003, ApJ. 582, 793

Goy, G. 1980, A&A Sup. Ser 42, 91

Gum, C. S. 1955, MNRAS, 67, 155

Georgelin, Y. M., Russeil, D., Amram, P., Georgelin, Y. P., Marcelin, M., Parker, W. A.,
    & Viale, A.  2000, A&A, 357, 308

Gaensler, B. M., Madsen, G., Chatterjee, S., & Mao, S. A. 2008, Pub. Astro. Soc. Aus.,
    25, 184

Hoopes, C., Sembach, K. R., Howk, J. C., Savage, B. D., & Fullerton, A. W. 2002,  ApJ,
    569, 233

Howk, J. C., & Savage, B. D. 1999, ApJ, 517, 746

Huang, J. S., Songaila, A., & Cowie, L. 1994, ApJ, 450, 163

Ito, M., & Ikeuchi, S. 1988, PASJ, 40, 403

Jenkins, E. B. 1978, ApJ, 219, 845

____________. 2000, IAU Joint Discussion 11, 6

Kalberla, P.M.W., & Dedes, L. 2008, A&A, 487, 951

Keeney, B. A., Danforth, C. W.,  Stocke, J. T.,  Penton, S. V., Shull, J. M., & Sembach,

    K. R.  2006, ApJ, 646, 951

Lampton, M., Margon, B., & Bowyer, S. 1976, ApJ, 208, 177

Lehner, N., & Howk, J. C. 2007, MNRAS, 377, 687

Morton, D.C. 2003, ApJS, 149, 205

Lipman, K., & Pettini, M. 1995, ApJ, 442, 628

Pettini, M. , & West, K. W. 1982, ApJ, 260, 561

Reynolds, R.  1989, ApJ, 339, L29

Richter, P. et al. 2001, ApJ, 559, 318

Savage, B. D., Edgar, R. & Diplas, A. 1990, ApJ, 361, 107

Savage, B. D., & Lehner, N. 2006, ApJS, 162, 134

Savage, B. D., Meade, M. R., & Sembach, K. R. 2001, ApJS, 136, 631

Savage, B. D., & Sembach, K. R. 1994, ApJ, 434, 145

Savage, B. D., Sembach, K. R., & Cardelli, J. A. 1994, ApJ, 420, 183

Savage, B. D., & Sembach, K. R. 1991, ApJ, 379, 245

Savage, B. D., & Sembach, K. R. 1996, ARA&A, 34, 279

Savage, B. D., Sembach, K. R., & Howk, J. C. 2001, ApJ, 547, 907

Savage, B. D., Sembach, K. R., & Lu, L. 1997, AJ, 113, 2158

Savage, B. D.,  Sembach, K. R., Wakker, B. et al. 2003, ApJS, 146, 125

Sharpless, S. 1957, PASP, 69, 239

Shelton, R. L. 1998, ApJ, 638, 206

_____________. 2006, ApJ, 504, 785

Shull, J. M., & Slavin, J. D. 1994, ApJ, 427, 784

Sembach, K. R., & Savage, B. D. 1992, ApJS, 83, 147

Sembach, K. R.,  Savage, B. D.,  & Tripp, T. M. 1997, ApJ, 480, 216

Thom, C., Peek, J.E.G., Putman, M. E. et al. 2008, ApJ, 684, 364

Wakker, B. P.  2001, ApJS, 136, 463

Wakker, B., Savage, B. D., Sembach, K. R.   et al. 2003, ApJS, 146, 1

Wakker, B., York, D. G., Howk, J. C. et al.   2007, ApJ, 670, L113

Wakker, B., York, D. G., Wilhelm, R. et al. 2008, ApJ, 672, 298

Van Woerden, H., & Wakker, B. P. in "High-Velocity Clouds", Astrophysics and Space
    Science Library, vol. 312, Kluwer-Dordrecht, p195.

Zsargó, J., Sembach, K. R., Howk, J. C. & Savage, B. D.  2003, ApJ,  586, 1019

APPENDIX

**Determining the Electron Scale Height from Pulsar Dispersion Measures**

Gaensler et al. (2008) have reported that the electron Galactic scale height based on the dispersion measures of pulsars is a factor of ~2 times larger than previously determined. The Gaensler et al. analysis techniques illustrate a number of the problems associated with obtaining reliable estimates of the |z| distribution of interstellar species when the gas distribution is irregular. Gaensler et al. collected electron dispersion measures, DM ($cm^{-3}$ pc), for pulsars with reliable distances estimated either from parallaxes or from the pulsar locations in globular clusters. For globular clusters with more than one pulsar they adopted the mean dispersion measure for the cluster and considered the mean as a single measurement in the scale height analysis. Their final sample included 51 lines of sight into the Galactic disk and halo. They constructed plots of log(DMsin|b|) versus log|z| and considered various ways of fitting the observations in order to reduce the effects of the contamination of the lines of sight by pronounced Galactic H II regions in the disk. They decided to limit the fit to the 15 pulsars at relatively high latitude with |b| > 40º when determining their preferred scale height estimate. Since the dispersion measure errors were very small compared to the distance errors, Gaensler et al. performed the fit using |z| as the dependent variable. They obtained h = 1.89 (+0.12, - 0.25) kpc which is approximately a factor of two larger than previous estimates. See Table 1 for the other fit parameters. There are several problems with the Gaensler et al. fit to the pulsar dispersion measures. Five of their six data points with |z| > 1 kpc lie an average of ~0.08 dex (~20%) below the fitted curve (see the blue dashed curve and blue data points in Figure 1 of Gaensler et al. 2008). The sixth data point is on the fitted curve. This suggests a systematic problem associated with the fit process. By using |z| as the dependent variable, fits to the observations can not be made if any of the data points at large |z| violate the condition DMsin|b| (observation) > DMsin|b| (model). If that condition is violated, the difference between the observation and the model treating |z| as the dependent variable, |z| (observed) - |z| (model), is undefined (i.e. forbidden). This constraint creates a strong bias in the fitted curve causing the derived scale height to be too large. In addition, the reported fit $\chi_\nu^2 = 5.2$ is large implying the fit is of low quality even though an additional 10% error was added to the errors on |z| for each measurement. The large value of $\chi_\nu^2$ is most likely the result of the patchy nature of the ISM which should be built into the model or into the fitting procedure. The Gaensler et al. scale height fit is invalid.

We discussed our concerns about the new electron scale height with Dr. Bryan Gaensler. He provided us with the values of DM and distance for the 15 pulsars with |b| > 40°. We used our software for estimating the electron scale height by considering DMsin|b| to be the dependent variable and followed the fitting procedures used for the absorption line measurements presented in this paper. The logarithmic errors for DM, distance, and the patchness parameter were combined in quadrature to produce a total error including the patchness $\sigma$ (total) = $[\sigma_{DM}^2 + \sigma_d^2 + \sigma_P]^{1/2}$. The distance errors for pulsars in the globular clusters were not included in the total error because at their large z distances, DMsin|b| achieves its maximum value. Again $\sigma_P$ was adjusted in order to achieve an acceptable fit with $\chi_\nu^2 = 1.0$. We obtain for $\sigma_P = 0.06$ dex, h = 1.41 (+0.26, -0.21) kpc and log(DMsin|b|) = 1.34±0.04 which corresponds to log[N(e)sin|b|] = 19.83±0.04. The other fit parameters are listed in Table 1. The fit method closely

parallels the approach we have taken in this paper.   We believe this second estimate of the electron scale height is more reliable than that given in Gaensler et al.  because it addresses the basic fact that the medium being studied is inhomogeneous even for relatively high latitude directions.  By using DMsin|b| as the dependent variable, the fit process does not forbid fits when DMsin|b| (observed) exceeds DMsin|b| (model) at large |z|.

The revised electron scale height of 1.41 (+0.26, -0.21)  kpc from the pulsar dispersion measures is in reasonable agreement with h = 0.90 (+0.62, -0.33) kpc derived by us for Al III which should also trace the WIM.   The Al III scale height fit required $\sigma_P$ = 0.256 dex to achieve $\chi_\nu^2$ = 1.0 which is considerably larger than the value 0.07 dex required for the pulsar dispersion measures.

The revised electron scale height of h = 1.41 (+0.26, -0.21)  kpc may contain substantial systematic errors because of the small size of the data sample and the biased sampling of the high latitude sky.  Ten of the 15 measurements at  |b| > 40° lie in the north Galactic sky. Five measurements have b > 70°.  Five measurements lie in the southern Galactic sky with b between -40° and -60°.   The high latitude north and south Galactic sky is asymmetrical in the distribution of many species including H I,  Si IV, C IV, and O VI.   In fact for the analysis of Si IV, C IV, and O VI in this paper,  the northern Galactic sky with |b| > 45° was excluded from the scale height derivation because all three species exhibited an excess of ~0.2 dex or 46% in the values of logN(ion)sin|b| toward extragalactic objects compared with the rest of the sample.  In the case of  H I,  the asymmetry is in the opposite sense.  Nine AGNs with b > 45° have <log[N(H I)sin|b|]> = 20.16 ± 0.19(STDEV) while nine with b < -45° have  <log[N(H

I)sin|b|]> = 20.25 ± 0.15(STDEV).   The north/south asymmetry in H I is  0.09 dex  or

23% when averaged over nine sightlines in the north and in the south .  Possible origins of the high ionization column density asymmetries are discussed in Savage et al. (2003).

A north/south Galactic asymmetry in the WIM is confirmed from a study of Fe III

and S III absorption with FUSE toward AGNs (Wakker et al. in preparation). Fe III is

produced and destroyed in photoionized gas with photon energies between 16.19 and

30.65 eV.  For S III the range is  23.34 and 34.79 eV.   While the Fe III observations are

affected by the presence of Fe in dust,  sulfur is mostly found in the gas phase (Savage &

Sembach 1996).  The existence of dust grain cores  in the WHIM have been revealed via studies of S III and Al III (Howk & Savage 1999). While some Fe III and S III may be produced in collisionally ionized gas, photoionization is likely the primary production mechanism for these ions.

The  column density of Fe III in the Galactic thick disk was obtained from FUSE

observations of the unsaturated line at 1122.208 Å with an oscillator strength f = 0.0788.

The S III column density is from FUSE observations of the unsaturated 1012.502 Å line with f = 0.0355.   These f values are from Morton (2003).  For both ions  the column densities are from the AOD method and refer to gas absorbing between ~±100 km s$^{-1}$.

The objects are drawn from the AGN sample listed in Table 2.  We list in Table 5 the

relevant averages of log[N(X)sin|b|] for lines of sight to AGNs in directions in the north

Galactic sky with  b >40º and in the south with b <-40º.  The table lists the object number

and values of <log[N(X)sin|b|]>±Stdev for Fe III  and  S III.   Although there is a

considerable dispersion in the results as revealed in the standard deviations which range from 0.16 to 0.20 dex, both ions reveal an excess in the logarithmic average for b > 40º compared to b <-40°.  For Fe III the excess is 0.20 dex or 58%.  For S III the excess is 0.15 dex or 41%.  The larger excess for Fe III may be related to differences in the level of dust processing in the north versus south Galactic polar regions.

Given the known north/south Galactic asymmetries, it would be highly desirable to obtain dispersion measures toward a sample of distant pulsars that covers the south Galactic sky at |b| <-40°.  Given the strong asymmetry in Fe III and S III which likely trace photoionized gas in the WIM,  the electron dispersion measures are expected to show an excess of ~0.15 to 0.2 dex in the north compared to the south.  The Gaensler et al. (2008) sample contains 4 globular clusters with b > 40° and 2 with b<-40°. The 4 globular clusters with b > 40° have $<DM\sin|b|>$ = 22.7 pc cm$^{-3}$ while the 2 with b <-40° have $<DM\sin|b|>$ = 17.8 pc cm$^{-3}$.  Although the sample is small, the existing observations imply a 27.5% or 0.11 dex excess of DMsin|b| in the north Galactic polar region.  The derivation of the scale height from the pulsar data should allow for the presence of the north/south asymmetry.  If the electrons behave like the transition temperature ions, the origin of the asymmetry is an excess in the north Galactic sky for b>40° compared to the rest of the sky with -90 < b < 40°

**Table 1. Galactic Exponential Scale Heights from the Literature based on the N(X)sin|b| vs |z|  Method**

| Species | N galactic | N extra-galactic | $\sigma_p$ (dex) | $n_0 \pm \sigma$ ($cm^{-3}$) | $\log(n_0 h) \pm \sigma$ (dex) | $h \pm \sigma$ (kpc) | Source/ comment |
|---|---|---|---|---|---|---|---|
| H I | 28 | 14 | 0.11 | 0.16±0.01 | 20.28±0.04 | 0.39±0.03 | 1 |
| Si IV | 30 | 5 | 0.31 | (2.3±0.2)x$10^{-9}$ | 13.56±0.06 | 5.1±0.7 | 1 |
| C IV | 30 | 13 | 0.30 | (9.2±0.8)x$10^{-9}$ | 14.08±0.07 | 4.4±0.6 | 1 |
| N V | 25 | 12 | … | (2.0±0.5)x$10^{-9}$ | 13.40±0.09 | 3.9±1.4 | 1 |
| O VI | 130 | 100 | 0.25 | 1.7x$10^{-8}$ | 14.09 | 2.3 | 2 |
| O VI | 180 | 40 | 0.25 | (1.33±0.15)x$10^{-8}$ | 14.28 | 4.6±1.0 | 3 |
| O VI | 180 | 30 | 0.28 | (1.34±0.17)x$10^{-8}$ | 14.12 | 3.2±0.8 | 3 |
| Al III | 29 | 2 | 0.20 | 2.2x$10^{-9}$ | 12.84±0.09 | 1.02(+0.36,-0.24) | 4 |
| $e^-$ | 41 | 0 | 0.22 | 0.036 | 20.00±0.09 | 0.89(+0.25, -0.18) | 5 |
| $e^-$ | 15 | 0 | … | 0.014±0.001 | 19.90±0.02 | 1.89(+0.12, -0.25) | 6 |
| $e^-$ | 15 | 0 | 0.07 | 0.016 | 19.83±0.04 | 1.41(+0.26, -0.21) | 7 |
| Ti II | 112 | 4 | 0.29 | 2.6x$10^{-10}$ | 12.04 | 1.35 | 8 |

Source/comment. (1) Savage et al. (1997); Results for Si IV are based on a small number of extragalactic objects.  The scale height of 3.9±1.4 kpc listed for N V properly allows for the fact 4 of the 12 extragalactic N V measurements are upper limits.   (2) Savage et al. (2003); Result excludes lines of sight through the north Galactic polar region which has a 0.25 dex excess of O VI. The preliminary value of $n_0$=1.7x$10^{-8}$ $cm^{-8}$ from the FUSE disk survey (Jenkins 2000) was adopted. (3) Bowen et al. (2008);  The larger O VI scale height is for the northern Milky Way with b>20º, the smaller is for the southern Milky Way with b <-20º. (4) Savage, Edgar & Diplas (1990); Al III traces the warm ionized medium.  The result is affected by the strong incorporation of Al into dust.   (5) Savage et al. (1990); Result is based on pulsar dispersion measures of the electron column density from Reynolds (1989) combined with values for M13 and M53 using the same analysis method as for the other UV and optical measurements listed in the Table. (6) This electron scale height from Gaensler et al. (2008) derived for 15 lines of sight to pulsars with |b| > 40° does not allow for the irregular (patchy) nature of the WIM even though $\chi_\nu^2 = 5.2$ implies the fit is poor. The fit considered |z| as the dependent variable which introduces a systematic fitting error as discussed in the appendix. (7) The values listed are for a reduced chi squared fit to the 15 lines of sight to pulsars with |b| > 40° from Gaensler et al. (2008) but including a patchiness parameter added in quadrature to the electron dispersion measure error and distance errors until $\chi_\nu^2 = 1.0$. Further details are provided in the Appendix. (8) Lipman & Pettini (1995); Ti II traces the warm neutral medium. The result is affected by the strong incorporation of Ti into dust particularly in the CNM.

## TABLE 2. Column Densities

| Object | MK | l (°) | b (°) | d (kpc) | σ (dex) | H II | [R] | logN(H I) (dex) | logN(Al III) (dex) | logN(Si IV) (dex) | logN(C IV) (dex) | logN(O VI) (dex) | source Al III, Si IV & C IV | source O VI |
|---|---|---|---|---|---|---|---|---|---|---|---|---|---|---|
| (1) | (2) | (3) | (4) | (5) | (6) | (7) | (8) | (9) | (10) | (11) | (12) | (13) | (14) | (15) |
| HD 167402 | BO.5Ib/BOII | 2.27 | -6.39 | 9.0 | 0.17 | … | 0 | 20.99±0.13 | … | 13.58±0.06 | 14.20±0.03 | 15.00±0.06 | 1 | 2 |
| HD 168941 | O9.5II-III | 5.82 | -6.31 | 6.8 | 0.17 | … | 0 | 21.11±0.09 | 12.94±0.07 | 13.67±0.16 | 14.39±0.10 | 15.03±0.03 | 1 | 2 |
| HD 165052 | O6.5Vnf | 6.12 | -1.48 | 1.9 | 0.08 | S25 | 2 | 21.36±0.10 | 13.40±0.06 | 13.66±0.33 | 13.96±0.06 | 13.89±0.06 | 1 | 2 |
| HDE 31 5021 | B2IVn | 6.12 | -1.33 | 1.4 | 0.16 | … | 1 | 21.28±0.10 | 13.62±0.14 | 13.71±0.16 | 13.99±0.11 | 13.98±0.10 | 1 | 2 |
| HD 163892 | O9IVn | 7.20 | 0.60 | 1.6 | 0.12 | … | 0 | 21.32±0.10 | 12.82±0.05 | <13.04 | 13.33±0.13 | 13.42±0.36 | 1 | 2 |
| HD 167771 | O7III:nf | 12.70 | -1.13 | 2.3 | 0.08 | S41 | 2 | 21.08±0.12 | 13.10±0.06 | >12.95 | 13.20±0.07 | 14.23±0.07 | 1 | 2 |
| HD 157857 | O6.5IIIf | 12.97 | 13.31 | 3.1 | 0.07 | … | 0 | 21.30±0.09 | 13.08±0.09 | 13.55±0.08 | 14.28±0.06 | 14.20±0.07 | 1 | 2 |
| HD 175876 | O6.5III | 15.28 | -10.58 | 2.7 | 0.07 | … | 0 | 21.04±0.11 | 13.09±0.02 | 13.22±0.15 | 13.51±0.15 | 14.14±0.12 | 1 | 3 |
| HD 175754 | O8IIf | 16.39 | -9.91 | 2.8 | 0.09 | … | 0 | 21.04±0.10 | 13.17±0.01 | 13.65±0.01 | 14.14±0.01 | 14.06±0.12 | 1 | 2 |
| HD 177989 | B0III | 17.81 | -11.88 | 6.0 | 0.15 | a | 0 | 20.95±0.09 | 13.39±0.03 | 13.80±0.01 | 14.47±0.08 | 14.31±0.06 | 6 | 3 |
| HD 178487 | B0.5Ib | 25.79 | -8.56 | 5.7 | 0.11 | … | 0 | 21.15±0.10 | … | 13.61±0.10 | 14.09±0.04 | 14.77±0.11 | 1 | 2 |
| vZ 1128 | PAGB | 42.50 | 78.69 | 10.0 | 0.04 | … | 1 | 19.98±0.03 | … | 13.83±0.05 | 14.44±0.06 | 14.48±0.02 | 13 | 13 |
| HD 187282 | WN4(h)+OB? | 55.62 | -3.79 | 4.5 | 0.09 | … | 0 | 21.38±0.15 | 13.22±0.07 | >13.78 | 14.59±0.21 | 14.39±0.10 | 1 | 2 |
| HD 191877 | B1Ib | 61.57 | -6.45 | 2.3 | 0.15 | … | 0 | 20.89±0.10 | … | >12.91 | 13.51±0.04 | 13.75±0.17 | 1 | 2 |
| HDE 33 2407 | B0.5III | 64.28 | 3.11 | 2.6 | 0.14 | … | 0 | 21.24±0.14 | 13.26±0.04 | 13.22±0.14 | 13.74±0.31 | 14.32±0.08 | 1 | 2 |
| HD 187459 | B0.5Ib | 68.81 | 3.85 | 2.2 | 0.11 | … | 0 | … | >12.49 | 12.34±0.14 | <13.02 | 13.30±0.30 | 1 | 2 |
| HD 190429 | O4 | 72.59 | 2.61 | 2.9 | 0.13 | … | 0 | … | 13.57±0.01 | >14.22 | >14.74 | 14.15±0.07 | 1 | 2 |
| HD 190918 | WN4+O9.7Iab | 72.65 | 2.06 | 2.1 | 0.09 | … | 2 | 21.40±0.10 | 13.70±0.03 | >14.18 | >14.75 | 14.09±0.05 | 1 | 2 |
| HD 191765 | WN6 | 73.45 | 1.55 | 1.7 | 0.09 | … | 1 | 21.48±0.14 | 13.98±0.04 | >14.23 | >14.80 | 14.08±0.10 | 1 | 2 |
| HD 201345 | O9V | 78.44 | -9.54 | 2.2 | 0.09 | … | 0 | 20.87±0.10 | 12.98±0.07 | >13.35 | 13.30±0.13 | 13.87±0.24 | 1 | 2 |
| HD 219188 | B0.5IIIn | 83.03 | -50.17 | 2.1 | 0.14 | … | 0 | 20.75±0.09 | 12.41±0.04 | 13.09±0.06 | 13.81±0.07 | 13.97±0.06 | 1 | 3 |
| HD 199579 | O6Vf | 85.70 | -0.30 | 1.4 | 0.06 | S117 | 1 | 21.04±0.11 | 13.33±0.06 | >13.5 | 13.51±0.10 | <13.38 | 1 | 2 |
| HD 195965 | B0V | 85.71 | 5.00 | 1.1 | 0.09 | … | 1 | 20.90±0.09 | 12.60±0.13 | 12.78±0.09 | >13.31 | 13.23±0.13 | 1 | 2 |
| HD 210809 | O9Iab | 99.85 | -3.13 | 4.3 | 0.10 | … | 0 | 21.25±0.07 | 13.26±0.09 | 12.62±0.11 | 13.09±0.09 | 14.08±0.10 | 1 | 2 |
| HD 212044 | B1Vnnpe | 100.64 | -4.35 | 0.9 | 0.17 | … | 0 | 21.00±0.07 | 12.94±0.05 | <13.01 | <13.25 | 13.82±0.23 | 1 | 2 |
| HD 210839 | O6Infp | 103.83 | 2.61 | 1.1 | 0.14 | … | 0 | 21.15±0.12 | 12.81±0.36 | 13.70±0.21 | <12.87 | <13.19 | 1 | 2 |
| HD 218915 | O9.5Iabe | 108.06 | -6.89 | 5.0 | 0.10 | … | 2 | 21.11±0.13 | … | 13.10±0.11 | 13.67±0.05 | 14.23±0.23 | 1 | 2 |
| HD 224151 | B0.5II | 115.44 | -4.64 | 1.4 | 0.14 | … | 0 | 21.32±0.10 | 13.29±0.51 | <12.99 | <13.40 | 13.61±0.12 | 1 | 2 |
| HD 5005A | O6.5Vf | 123.13 | -6.24 | 2.9 | 0.02 | S184 | 1 | 21.40±0.10 | 13.12±0.06 | >13.69 | 13.88±0.17 | <13.38 | 1 | 2 |
| HD 14434 | O5.5Vnfp | 135.08 | -3.82 | 3.5 | 0.07 | … | 0 | 21.45±0.08 | 13.04±0.04 | <12.99 | 13.94±0.21 | <14.12 | 1 | 2 |
| HD 34656 | O7IIf | 170.04 | 0.27 | 2.2 | 0.08 | … | 0 | … | 13.08±0.04 | >12.73 | 13.07±0.20 | 13.97±0.07 | 1 | 2 |
| BD +38 2182 | B3V | 182.16 | 62.21 | 4.5 | 0.07 | … | 0 | … | 12.86±0.05 | 13.55±0.08 | 13.97±0.04 | 14.10±0.07 | 1 | 3 |
| HD 42088 | O6.5V | 190.04 | 0.48 | 2.0 | 0.08 | S252 | 1 | 21.15±0.08 | 13.44±0.07 | 14.07±0.35 | 13.95±0.07 | 13.57±0.23 | 1,PF | 2 |
| HD 39680 | O6.5:nep | 194.08 | -5.88 | 2.6 | 0.07 | ... | 0 | 21.30±0.08 | 12.91±0.08 | <12.53 | <13.07 | 13.69±0.29 | 1 | 2 |
| HD 45314 | O9pe | 196.96 | 1.52 | 2.1 | 0.33 | ... | 3 | 21.04±0.09 | 12.48±0.05 | >12.87 | 13.72±0.04 | 14.44±0.03 | 1 | 2 |
| HD 42401 | B2V | 197.64 | -3.33 | 0.8 | 0.13 | .. | 0 | ... | ... | <13.31 | <13.30 | <13.68 | 1 | 2 |
| HD 47417 | B0IV | 205.35 | 0.35 | 1.3 | 0.13 | ... | 3 | 21.13±0.09 | 13.33±0.08 | >13.80 | 13.95±0.07 | 14.06±0.15 | 1 | 2 |
| HD 46150 | O5Vf | 206.31 | -2.07 | 1.7 | 0.06 | S275 | 2 | 21.25±0.12 | 13.30±0.05 | >13.67 | 13.77±0.03 | 13.78±0.10 | 1 | 2 |
| HD 47360 | B0.5V | 207.33 | -0.79 | 1.5 | 0.09 | ... | 3 | 21.46±0.08 | 13.16±0.03 | 13.37±0.17 | 14.01±0.07 | 13.86±0.25 | 1 | 2 |
| HD 18100 | B1V | 217.93 | -62.73 | 3.1 | 0.07 | ... | 1 | 20.14±0.07 | 12.71±0.05 | 13.10±0.04 | 13.58±0.02 | 13.77±0.12 | 4 | 3 |
| HD 65079 | B2Vne | 217.87 | 15.93 | 1.0 | 0.14 | ... | 0 | ... | >12.84 | 13.23±0.29 | 13.89±0.32 | 13.65±0.15 | 1 | 2 |
| HD 61347 | O9Ib | 230.60 | 3.79 | 4.4 | 0.09 | ... | 0 | 21.53±0.13 | 12.93: | 13.17±0.13 | 14.00±0.41 | 14.30±0.14 | 1 | 2 |
| HD 58510 | B1Ib-II | 235.52 | -2.47 | 2.5 | 0.24 | ... | 0 | 21.28±0.12 | 12.99±0.10 | >13.34 | 13.68: | 14.15±0.12 | 1 | 2 |
| HD 38666 | O9.5V | 237.29 | -27.10 | 0.8 | 0.09 | ... | 0 | 19.84±0.08 | 12.01±0.05 | 12.16±0.05 | 12.88±0.02 | 13.82±0.02 | 9,12 | 9 |
| HD 52463 | B3 V | 238.90 | -10.52 | 0.9 | 0.15 | ... | 0 | ... | 13.02±0.21 | 12.75±0.16 | 13.10±016 | <13.55 | 1 | 2 |
| HD 63005 | O6Vf | 242.47 | -0.93 | 5.4 | 0.07 | ... | 0 | ... | 12.94±0.04 | ... | 14.03±0.10 | 14.27±0.06 | 1 | 2 |
| HD 66788 | O9V | 245.43 | 2.05 | 4.3 | 0.09 | ... | 0 | ... | 13.44±0.02 | >14.28 | 14.03±0.05 | 14.36±0.06 | 1 | 2 |
| HD 69106 | B0.5IVnn | 254.52 | -1.33 | 1.5 | 0.13 | ... | 0 | 21.08±0.06 | 13.09±0.01 | 13.23±0.07 | 13.83±0.09 | 14.31±0.06 | 1 | 2 |
| HD 100340 | B1V | 258.85 | 61.23 | 3.0 | 0.03 | ... | 0 | 20.47±0.06 | 12.88±0.07 | 13.31±0.09 | 13.83±0.02 | 14.22±0.14 | 4 | 3 |
| HD 64760 | B0.5Ib | 262.06 | -10.42 | 1.1 | 0.11 | b | 0 | 20.22±0.08 | 13.69±0.11 | 13.44±0.06 | 13.96±0.04 | 14.08±0.02 | 8 | 7 |
| HD 68273 | O9I+WC8 | 262.80 | -7.69 | 0.3 | 0.06 | b | | 19.74±0.06 | 13.21±0.17 | 13.05±0.03 | 13.18±0.03 | 13.71±0.02 | 10,8 | 10 |
| HD 42933 | B0.5IV | 263.30 | -27.68 | 0.8 | 0.13 | | 2 | 20.23±0.08 | ... | <12.90 | 13.03±0.05 | 13.59±0.06 | 11 | 7 |
| HD 74920 | O8 | 265.29 | -1.95 | 3.6 | 0.29 | c | 3 | 21.15±0.09 | 13.39±0.01 | 14.51±0.12 | 13.78±0.09 | 14.29±0.05 | 1 | 2 |
| HD 74711 | B1III | 265.74 | -2.61 | 1.2 | 0.15 | c | 3 | ... | 13.15±0.02 | 13.46±0.06 | 14.28±0.17 | 13.78±0.14 | 1 | 2 |
| HD 89137 | O9.7IIInp | 279.70 | 4.45 | 3.1 | 0.14 | ... | 0 | ... | 12.94±0.13 | 13.00±0.12 | 13.65±0.15 | 13.96±0.17 | 1 | 2 |
| HD 90087 | O9.5III | 285.16 | -2.13 | 2.8 | 0.11 | ... | 0 | 21.15±0.06 | 12.81±0.03 | <13.13 | 13.43±0.09 | 14.06±0.06 | 1 | 2 |
| HD 88115 | B1.5IIn | 285.32 | -5.53 | 3.7 | 0.21 | ... | 0 | 21.02±0.11 | 13.30±0.03 | <12.89 | 13.74±0.34 | 14.15±0.30 | 1 | 2 |
| HD 91651 | O9V:n | 286.55 | -1.72 | 2.8 | 0.09 | ... | 0 | 21.15±0.06 | 13.69±0.05 | 13.89±0.07 | 14.03±0.05 | 14.50±0.08 | 1 | 2 |
| HD 91597 | B1IIIne | 286.86 | -2.37 | 3.9 | 0.14 | ... | 1 | 21.40±0.06 | >13.82 | >13.84 | >14.33 | 14.49±0.02 | 1 | 2 |
| HD 93129A | O2If | 287.41 | -0.57 | 2.8 | 0.09 | d | 2 | ... | 13.87±0.02 | 14.42±0.23 | 14.74±0.05 | 14.67±0.06 | 1 | 2 |
| HD 93250 | O3Vf | 287.51 | -0.54 | 2.3 | 0.05 | d | 2 | 21.39±0.15 | 14.05±0.07 | >14.36 | 14.60±0.05 | 14.63±0.03 | 1 | 2 |
| HD 93205 | O3Vf+ | 287.57 | -0.71 | 3.3 | 0.06 | d | 2 | 21.33±0.10 | 13.96±0.04 | 14.18±0.31 | 14.20±0.06 | 14.41±0.14 | 1 | 2 |
| CPD -59 2603 | O7Vf | 287.59 | -0.69 | 3.5 | 0.08 | d | 2 | 21.46±0.07 | 13.86±0.10 | 14.24±0.15 | 14.67±0.10 | 14.16±0.13 | 1 | 2 |
| HDE 30 3308 | O3Vf | 287.60 | -0.61 | 3.8 | 0.06 | d | 2 | 21.45±0.09 | 13.89±0.04 | 14.30±0.17 | 14.55±0.06 | 14.76±0.07 | 1 | 2 |
| HD 93146 | O6.5Vf | 287.67 | -1.05 | 3.5 | 0.07 | d | 1 | 21.18±0.09 | 14.05±0.05 | >14.40 | 14.70±0.09 | 14.79±0.02 | 1 | 2 |
| HD 93206 | O9.7Ib:n | 287.67 | -0.94 | 2.6 | 0.08 | d | 2 | 21.34±0.11 | 13.92±0.02 | >14.34 | 14.57±0.03 | 14.81±0.02 | 1 | 2 |
| HD 93222 | O7IIIf | 287.74 | -1.02 | 3.6 | 0.09 | d | 1 | ... | 13.82±0.05 | >14.26 | 14.34±0.02 | 14.71±0.05 | 1 | 2 |
| HD 93843 | O5IIIf | 288.24 | -0.90 | 3.5 | 0.08 | ... | 0 | 21.33±0.08 | 13.48±0.05 | 13.50±0.02 | 13.98±0.05 | 14.38±0.05 | 1 | 2 |
| HD 94493 | B1Ib | 289.02 | -1.18 | 3.4 | 0.15 | ... | 1 | 21.11±0.09 | ... | 13.06±0.09 | 13.58±0.03 | 14.02±0.08 | 1 | 2 |
| HD 96917 | O8.5Ibf | 289.29 | 3.06 | 2.9 | 0.09 | ... | 0 | 21.23±0.11 | 13.39±0.06 | 13.57±0.15 | 14.29±0.06 | 14.61±0.21 | 1 | 2 |
| HD 96670 | O8Ibp | 290.20 | 0.40 | 3.8 | 0.17 | ... | 1 | ... | 13.68±0.13 | 13.36±0.57 | 13.68±0.13 | 13.95±0.32 | 1 | 2 |

| HD 96715 | O4Vf | 290.27 | 0.33 | 3.8 | 0.07 | ... | 1 | 21.20±0.16 | 13.07±0.12 | 13.50±0.04 | 14.07±0.15 | 14.44±0.13 | 1 | 2 |
|---|---|---|---|---|---|---|---|---|---|---|---|---|---|---|
| HD 99890 | B0IIIn | 291.75 | 4.43 | 3.5 | 0.15 | ... | 0 | 20.93±0.13 | 13.48±0.08 | 13.48±0.13 | 14.07±0.06 | 14.17±0.07 | 1 | 2 |
| HD 100276 | B0.5Ib | 293.31 | 0.77 | 3.2 | 0.12 | ... | 0 | 21.19±0.09 | ... | 13.03±0.04 | 13.65±0.04 | 13.48±0.27 | 1 | 2 |
| HD 99857 | B0.5Ib | 294.78 | -4.94 | 3.5 | 0.11 | ... | 1 | 21.31±0.12 | 13.26±0.33 | ... | 13.79±0.09 | 13.75±0.17 | 1 | 2 |
| HD 101131 | O6Vf | 294.78 | -1.62 | 2.0 | 0.07 | R62[e] | 2 | ... | 13.71±0.04 | 13.91±0.09 | 13.83±0.03 | 14.16±0.03 | 1 | 2 |
| HD 101190 | O6Vf | 294.78 | -1.49 | 2.1 | 0.08 | R62[e] | 2 | 21.15±0.11 | 13.28±0.06 | 13.73±0.32 | >13.98 | 14.31±0.02 | 1 | 2 |
| HDE 308813 | O9.5V | 294.80 | -1.61 | 3.1 | 0.09 | R62[e] | 2 | 21.15±0.10 | 13.43±0.04 | >13.83 | 14.00±0.10 | 14.20±0.03 | 1 | 2 |
| HD 100213 | O8.5Vn | 284.81 | -4.14 | 2.6 | 0.08 | ... | 0 | 21.18±0.07 | 13.10±0.01 | 12.99±0.04 | 13.79±0.07 | 14.27±0.08 | 1 | 2 |
| HD 101205 | O7IIInf | 294.85 | -1.65 | 2.4 | 0.08 | R62[e] | 2 | 21.20±0.07 | 13.69±0.04 | 13.96±0.20 | 14.05±0.06 | 14.42±0.07 | 1 | 2 |
| HD 101298 | O6Vf | 294.94 | -1.69 | 2.8 | 0.08 | R62[e] | 2 | 21.26±0.11 | >13.63 | 13.83±0.17 | 13.90±0.02 | 14.26±0.02 | 1 | 2 |
| HD 101413 | O8V | 295.03 | -1.71 | 2.4 | 0.08 | R62[e] | 1 | 21.23±0.13 | 13.52±0.06 | 14.00±0.21 | 13.94±0.09 | 14.28±0.05 | 1 | 2 |
| HD 101436 | O6.5V | 295.04 | -1.71 | 2.2 | 0.08 | R62[e] | 1 | 21.23±0.08 | 13.81±0.08 | 13.91±0.12 | 13.99±0.07 | 14.26±0.05 | 1 | 2 |
| HD 102552 | B1IIIn | 295.21 | 1.35 | 3.9 | 0.15 | ... | 0 | ... | 13.16±0.05 | 13.27±0.13 | 13.24±0.23 | 13.61±0.18 | 1,PF | 2 |
| HD 103779 | B0.5Iab | 296.85 | -1.02 | 4.3 | 0.11 | ... | 1 | 21.16±0.10 | ... | 13.18±0.06 | 13.80±0.03 | 13.47±0.20 | 1 | 2 |
| HD 104705 | B0Ib | 297.46 | -0.34 | 5.0 | 0.09 | ... | 1 | 21.11±0.07 | 13.21±0.02 | 13.64±0.10 | 14.63±0.19 | 14.05±0.11 | 1 | 2 |
| JL 212 | B2V | 303.63 | -61.03 | 2.3 | 0.07 | ... | 0 | ... | 12.83±0.05 | >13.24 | 13.50±0.04 | 14.21±0.07 | 1 | 3 |
| HD 112244 | O8.5Iabf | 303.55 | 6.03 | 1.7 | 0.09 | ... | 0 | 21.08±0.08 | ... | 13.23±0.05 | 14.00±0.06 | 13.86±0.24 | 11 | 7 |
| HD 116852 | O9III | 304.88 | -16.13 | 4.6 | 0.12 | ... | 0 | 20.96±0.08 | 13.27±0.06 | 13.60±0.02 | 14.08±0.03 | 14.30±0.02 | 5 | 3 |
| HD 115071 | B0.5V | 305.77 | 0.15 | 1.2 | 0.12 | ... | 0 | 21.38±0.10 | 13.51±0.01 | 13.13±0.35 | >13.83 | 14.10±0.08 | 1 | 2 |
| HD 116781 | B0IIIne | 307.05 | -0.06 | 2.2 | 0.14 | ... | 0 | ... | 12.87±0.18 | <13.38 | <13.41 | 14.05±0.16 | 1 | 2 |
| HD 116538 | B2IVn | 308.23 | 10.68 | 1.3 | 0.15 | ... | 0 | ... | 13.05±0.14 | <12.76 | 13.29±0.07 | 13.38±0.14 | 1 | 2 |
| HD 124314 | O6Vnf | 312.67 | -0.42 | 1.4 | 0.08 | ... | 3 | 21.34±0.10 | 13.28±0.19 | 13.43±0.09 | 14.20±0.04 | 13.76±0.11 | 1 | 2 |
| HD 116658 | B1V | 316.11 | 50.84 | 0.1 | 0.05 | ... | 0 | ... | 12.26±0.18 | 13.18±0.06 | 13.43±0.07 | 11 | 7 |
| HD 124979 | O8Vf | 316.41 | 9.08 | 2.8 | 0.08 | ... | 0 | 21.30±0.11 | 13.22±0.03 | 12.91±0.14 | 13.44±0.11 | 14.08±0.07 | 1 | 2 |
| HD 148422 | B1Ia | 329.92 | -5.60 | 10.0 | 0.09 | ... | 1 | 21.15±0.12 | ... | 13.63±0.26 | 13.92±0.18 | 14.10±0.14 | 1 | 2 |
| HD 150898 | B0.5Ia | 329.98 | -8.48 | 2.8 | 0.09 | ... | 0 | 20.90±0.12 | ... | 13.82±0.01 | 14.17±0.03 | 13.84±0.22 | 11 | 7 |
| HD 121968 | B1V | 333.97 | 55.84 | 3.8 | 0.17 | ... | 0 | 20.60±0.08 | 12.91±0.02 | 13.67±0.23 | 14.21±0.10 | 13.97±0.06 | 1 | 3 |
| HD 151805 | B1II | 343.20 | 1.59 | 6.0 | 0.15 | ... | 1 | 21.36±0.11 | 13.40±0.12 | <13.28 | <13.57 | 14.06±0.12 | 1 | 2 |
| HD 151932 | WN7h | 343.23 | 1.43 | 2.0 | 0.09 | ... | 1 | 21.39±0.11 | 13.37±0.03 | 12.71±0.05 | 13.52±0.03 | 14.20±0.05 | 1 | 2 |
| HD 152248 | O7Ib:nfp | 343.47 | 1.18 | 2.2 | 0.09 | R113 | 2 | ... | 13.60±0.02 | 13.01±0.04 | 13.60±0.02 | 14.30±0.05 | 1 | 2 |
| HD 152233 | O6IIIfp | 343.48 | 1.22 | 2.3 | 0.08 | R113 | 2 | 21.29±0.10 | 13.34±0.01 | 13.20±0.62 | 13.67±0.19 | 14.32±0.02 | 1 | 2 |
| HD 152314 | O9.5III-IV | 343.52 | 1.14 | 2.6 | 0.19 | R113 | 2 | ... | 13.46±0.07 | <13.17 | 13.64±0.06 | 13.92±0.30 | 1 | 2 |
| HD 152623 | O7Vnf | 344.62 | 1.61 | 1.5 | 0.08 | R113[f] | 2 | 21.28±0.10 | 13.41±0.05 | 13.79±0.11 | 13.77±0.03 | 14.41±0.05 | 1 | 2 |
| HD 152723 | O6.5IIIf | 344.81 | 1.61 | 2.3 | 0.07 | R113[f] | 1 | 21.43±0.13 | 13.72±0.10 | 13.81±0.40 | 13.93±0.02 | 14.42±0.07 | 1 | 2 |
| HD 156292 | O9.5III | 345.35 | -3.08 | 1.7 | 0.13 | ... | 1 | ... | 13.01±0.15 | 13.26±0.09 | 13.95±0.07 | 14.30±0.08 | 1 | 2 |
| HD 153426 | O9II-III | 347.14 | 2.38 | 2.6 | 0.16 | S2[g] | 1 | 21.34±0.13 | 13.60±0.04 | 13.83±0.15 | 13.95±0.07 | 14.25±0.02 | 1 | 2 |
| HD 163758 | O6.5Iaf | 355.36 | -6.10 | 4.7 | 0.07 | ... | 1 | 21.23±0.18 | 13.72±0.05 | 13.67±0.03 | 14.02±0.02 | 14.49±0.08 | 1 | 2 |
| HD 177566 | PAGB | 355.55 | -20.42 | 1.1 | 0.18 | ... | 1 | 20.88±0.09 | 12.50±0.04 | <12.49 | 13.10±0.13 | 13.65±0.07 | 1 | 3 |
| TON_S210 | AGN | 224.97 | -83.16 | ... | ... | ... | 0 | 20.20±0.05 | ... | 13.62±0.06 | 14.43±0.06 | 14.57±0.02 | 13 | 13 |
| HE 0226-4110 | AGN | 253.94 | -65.77 | ... | ... | ... | 0 | 20.27±0.05 | ... | 13.41±0.04 | 13.95±0.07 | 14.00±0.09 | 13 | 13 |
| MRK 1044 | AGN | 179.69 | -60.48 | ... | ... | ... | 0 | ... | ... | ... | 13.87±0.07 | 14.26±0.13 | 13 | 13 |
| PKS 2155-304 | AGN | 17.73 | -52.25 | ... | ... | ... | 0 | 20.12±0.04 | ... | 13.36±0.04 | 14.06±0.02 | 14.31±0.02 | 13 | 13 |
| PHL 1811 | AGN | 47.47 | -44.81 | ... | ... | ... | 0 | 20.59±0.04 | ... | 13.61±0.02 | 14.30±0.02 | 14.39±0.03 | 13 | 13 |
| PKS 0405-12 | AGN | 204.93 | -41.76 | ... | ... | ... | 0 | 20.54±0.04 | ... | 13.43±0.04 | 14.08±0.05 | 13.97±0.05 | 13 | 13 |
| MRK 335 | AGN | 108.76 | -41.42 | ... | ... | ... | 0 | 20.43±0.04 | ... | 13.56±0.04 | 14.16±0.04 | 14.04±0.05 | 13 | 13 |
| NGC 1705 | AGN | 261.08 | -38.74 | ... | ... | ... | 0 | 20.25±0.04 | ... | 13.70±0.02 | 14.38±0.06 | 14.23±0.03 | 13 | 13 |
| MRK 509 | AGN | 35.97 | -29.86 | ... | ... | ... | 0 | 20.61±0.04 | ... | 13.89±0.02 | 14.51±0.03 | 14.68±0.01 | 13 | 13 |
| ESO 141-G55 | AGN | 338.18 | -26.71 | ... | ... | ... | 0 | 20.79±0.04 | ... | 13.91±0.07 | >14.57 | 14.53±0.03 | 13 | 13 |
| HS 0624+6907 | AGN | 145.71 | 23.35 | ... | ... | ... | 0 | 20.89±0.04 | ... | 13.72±0.05 | 14.02±0.04 | 14.47±0.05 | 13 | 13 |
| 3C 249.1 | AGN | 130.39 | 38.55 | ... | ... | ... | 0 | 20.45±0.04 | ... | 13.40±0.05 | 13.91±0.05 | 14.38±0.04 | 13 | 13 |
| MRK 876 | AGN | 98.27 | 40.38 | ... | ... | ... | 0 | 20.38±0.04 | ... | 13.73±0.02 | 14.28±0.03 | 14.48±0.02 | 13 | 13 |
| MRK 205 | AGN | 125.45 | 41.67 | ... | ... | ... | 0 | 20.48±0.04 | ... | 13.56±0.03 | 14.15±0.06 | 14.24±0.04 | 13 | 13 |
| NGC 3516 | AGN | 133.24 | 42.40 | ... | ... | ... | 0 | ... | ... | 13.34±0.09 | 13.91±0.06 | 14.42±0.09 | 13 | 13 |
| MRK 279 | AGN | 115.04 | 46.86 | ... | ... | ... | 0 | 20.27±0.04 | ... | 13.65±0.02 | 14.21±0.02 | 14.41±0.01 | 13 | 13 |
| PG 0953+414 | AGN | 179.79 | 51.71 | ... | ... | ... | 0 | 20.09±0.04 | ... | 13.34±0.07 | 13.95±0.05 | 14.39±0.03 | 13 | 13 |
| PG 1302-102 | AGN | 308.59 | 52.16 | ... | ... | ... | 0 | 20.50±0.04 | ... | 13.73±0.05 | 14.22±0.05 | 14.24±0.04 | 13 | 13 |
| MRK 1383 | AGN | 349.22 | 55.12 | ... | ... | ... | 0 | 20.40±0.04 | ... | 13.79±0.04 | 14.45±0.05 | 14.62±0.02 | 13 | 13 |
| PG 1259+593 | AGN | 120.56 | 58.05 | ... | ... | ... | 0 | 19.90±0.04 | ... | 13.35±0.04 | 13.79±0.06 | 14.15±0.03 | 13 | 13 |
| 3C 273.0 | AGN | 289.95 | 64.36 | ... | ... | ... | 0 | 20.22±0.04 | ... | 13.85±0.01 | 14.51±0.01 | 14.73±0.01 | 13 | 13 |
| PG 1116+215 | AGN | 223.36 | 68.21 | ... | ... | ... | 0 | 20.08±0.04 | ... | 13.70±0.02 | 14.21±0.03 | 14.23±0.03 | 13 | 13 |
| NGC 5548 | AGN | 31.96 | 70.50 | ... | ... | ... | 0 | 20.20±0.04 | ... | 13.82±0.02 | 14.45±0.03 | 14.48±0.06 | 13 | 13 |
| PG 1211+143 | AGN | 267.55 | 74.32 | ... | ... | ... | 0 | 20.42±0.04 | ... | 13.61±0.02 | 14.09±0.02 | 14.18±0.04 | 13 | 13 |
| NGC 3783 | AGN | 287.46 | 22.95 | ... | ... | ... | 0 | 20.98±0.04 | ... | 13.70±0.01 | 14.31±0.01 | 14.34±0.06 | 13 | 13 |
| SK -65 22 (LMC) | O6Iaf+ | 276.40 | -35.75 | 49.0 | ... | ... | 0 | ... | ... | 13.26±0.02 | 13.96±0.02 | 14.20±0.02 | 14 | 14 |
| SK -67 211(LMC) | O2III(f*) | 277.70 | -32.16 | 49.0 | ... | ... | 0 | ... | ... | 13.44±0.03 | 14.14±0.14 | 14.36±0.02 | 14 | 14 |
| SK -69 246(LMC) | WN6h | 278.38 | -31.66 | 49.0 | ... | ... | 0 | ... | ... | 13.37±0.02 | 13.92±0.03 | 14.11±0.08 | 14 | 14 |
| SK -71 45 (LMC) | O4-5III(f) | 281.86 | -32.02 | 49.0 | ... | ... | 0 | ... | ... | 13.52±0.03 | 14.10±0.02 | 14.34±0.02 | 14 | 14 |
| SN1987A (LMC) | SN | 279.70 | -31.94 | 49.0 | ... | ... | 0 | ... | 13.14±0.05 | 13.55±0.06 | 14.15±0.05 | ... | 15 | ... |
| HD 5980 (SMC) | WNp | 302.07 | -44.95 | 61.0 | ... | ... | 0 | ... | 12.97±0.04 | 13.82±0.11 | 14.35±0.04 | 14.25±0.06 | 15 | 16 |

Notes: [a] Line of sight samples gas in the Scutum Supershell (see Savage et al. 2001). [b] Line of sight samples gas in the Gum nebula. [c] Line of sight samples gas in the Vela SN remnant. [d] Line of sight samples gas in the Carina H II region. [e] Line of sight samples gas in Gum 42 which is also R62. [f] Line of sight samples gas in Gum 56 which is within R113. [g] Line of sight samples gas in Gum 57a which is within S2. Column density sources: (1)Bowen et al. 2008; (2) Savage, Meade & Sembach 2001; (3) Zsargó et al. 2003; (4) Savage & Sembach 1994; (5) Fox et al. 2003; (6) Savage, Sembach & Howk 2001; (7) Jenkins 1977; (8) Edgar & Savage 1992; (9) Brandt et al. 1999; (10) Fitzpatrick & Spitzer 1994; (11) Huang et al. 1994; (12) Howk & Savage 1999; (13) AGN results from Wakker, Savage & Sambach (in preparation); (14)LMC results from Lehner & Howk 2007; (15) Sembach & Savage 1992; (16) Hoopes et al. 2002.

**Table 3. log(n$_o$h), n$_o$ and h for the Restricted Object Sample[a]**

| Ion | N galactic | N extra-galactic | $\sigma_p$ (dex) | log(n$_o$h)$\pm\sigma$ (dex) | n$_o$ (cm$^{-3}$) | h$\pm\sigma$ (kpc) |
|------|------|------|------|------|------|------|
| H I | 52 | 14 | 0.172 | 20.31$\pm$0.05 | 0.276 | 0.24$\pm$0.06 |
| Al III | 52 | 2 | 0.256 | 12.75(+0.18, -0.16) | 2.02x10$^{-9}$ | 0.90(+0.62,-0.33) |
| Si IV | 44 | 21 | 0.266 | 13.36 (+0.09, -0.08) | 2.32x10$^{-9}$ | 3.2(+1.0,-0.6) |
| C IV | 55 | 21 | 0.273 | 13.95 (+0.08, -0.09) | 8.02x10$^{-9}$ | 3.6(+1.0,-0.8) |
| O VI | 64 | 21 | 0.233 | 14.12 (+0.07, -0.08) | 1.64x10$^{-8}$ | 2.6$\pm$0.5 |

[a] The restricted object sample excludes sight lines to stars in prominent H II regions, sight lines through known SN remnants, sight lines toward stars with strong foreground X-ray emission, and extragalactic sight lines toward the north Galactic pole with b > 45º along with vZ 1128. Limits are not included when performing the $\chi^2$ fit to the exponential scale height model which allows for a patchy distribution of the gas.

**Table 4. Sample Medians and Averages**

| Quantity | Sample[b] | Number | Median | Average | Stdev |
|---|---|---|---|---|---|
| log[N(O VI)sin\|b\|] | 1 | 20 | 14.07 | 14.10 | 0.20 |
| " | 2 | 29 | 14.15 | 14.17 | 0.22 |
| " | 3 | 9 | 14.27 | 14.31 | 0.20 |
| | | | | | |
| log[N(C IV)sin\|b\|] | 1 | 20 | 13.91 | 13.95 | 0.22 |
| " | 2 | 29 | 13.98 | 14.01 | 0.24 |
| " | 3 | 9 | 14.12 | 14.14 | 0.25 |
| | | | | | |
| log[N(Si IV)sin\|b\|] | 1 | 20 | 13.36 | 13.35 | 0.18 |
| " | 2 | 29 | 13.38 | 13.42 | 0.22 |
| " | 3 | 9 | 13.63 | 13.58 | 0.21 |
| | | | | | |
| log[N(Si IV)/N(C IV)] | 1 | 20 | -0.59 | -0.60 | 0.11 |
| " | 2 | 29 | -0.59 | -0.59 | 0.10 |
| " | 3 | 9 | -0.56 | -0.56 | 0.08 |
| " | 4 | 40 | -0.54 | -0.55 | 0.21 |
| | | | | | |
| log[N(C IV)/N(O VI)] | 1 | 20 | -0.17 | -0.15 | 0.17 |
| " | 2 | 29 | -0.17 | -0.16 | 0.18 |
| " | 3 | 9 | -0.17 | -0.17 | 0.20 |
| " | 4 | 53 | -0.32 | -0.29 | 0.35 |
| | | | | | |
| log[N(Si IV)/N(O VI)] | 1 | 20 | -0.75 | -0.74 | 0.19 |
| | 2 | 29 | -0.75 | -0.74 | 0.18 |
| | 3 | 9 | -0.78 | -0.73 | 0.18 |
| | 4 | 43 | -0.88 | -0.83 | 0.38 |

[b]Object samples: Limits  are not included in any of the following samples which utilize the column density measurements listed in Table 2.  (1) Extragalactic lines of sight with b<45º. (2) All extragalactic lines of sight. (3) Extragalactic lines of sight with b>45º. (4) The restricted galactic lines of sight.

**Table 5. North/South Asymmetry in the Galactic Distribution of Fe III and S III[1]**

| Ion | b | number | <log[N(X)sin|b|]>±stdev |
|-----|------|--------|------------------------|
| Fe III | <-40° | 9 | 13.94±0.16 |
| Fe III | >40° | 25 | 14.14±0.20 |
| S III | <-40° | 6 | 14.27±0.16 |
| S III | >40° | 20 | 14.42±0.17 |

[1]Averages of log[N(X)sin|b|] for gas in the Galactic thick disk toward AGNs observed by Wakker et al. (in preparation) reveal the strong north/south asymmetry in the distribution of these ions which likely trace photoionized gas in the WIM.

FIGURE CAPTIONS

FIG. 1a,- Sky Distribution of the Galactic and extragalactic lines of sight in an Aitoff projection with directions to survey stars shown as open circles and + symbols. Extragalactic objects are shown as open and filled stars. The + symbols  are for directions to stars contaminated by bright H II region emission or bright associated X-ray emission or associated foreground SNRs (see Table 2).  The filled stars are for directions toward extragalactic objects  with b > 45º which exhibit an excess of ~0.2 dex  in the projected (Nsin|b|) column densities of the high ions.  The symbol for the star vZ 1128  near the north Galactic pole at b = 78.69º and l = 42.50º is plotted with a + symbol since the absorption along the 10 kpc path also reveals an excess in Nsin|b|.

FIG.1b.-  Views of the object lines of sights from above the Galactic plane displayed in two different scales.   In the left panels the lines of sight to Galactic stars are shown as the solid lines.  The lines of sight to the extragalactic objects are shown with dashed lines that end when the  |z| distance is 8 kpc which corresponds to several times the high ion scale heights. In the right panels the circles show the projected  positions of  the Galactic stars. The star shows the position of the Sun.

FIG. 2.-  logN(X)sin|b| versus log|z| for X = H I, Al III, Si IV, C IV, O VI. Detections are plotted with the filled or open circles. Upper and lower limits are plotted with downward and upward pointing triangles, respectively.  The extragalactic measurements for objects beyond the Magellanic Clouds are ordered according to

Galactic latitude from -90° to +90°. The light lines shows the distribution of $\log(n_0 h)$ $+\log[1-\exp(-|z|/h)]$ versus $\log|z|$ representing a smooth exponential gas stratification with the mid-plane densities from Table 3 and scale heights of 0.1, 0.3, 1, 3 and 10 kpc. The best fit results and $\pm 1\sigma$ errors from Table 3 are shown with the thick and dashed lines.The large dispersion of the observed data points about such a simple model is the result of the irregular (patchy) distribution of the interstellar gas. The filled data points are for the restricted sample. The open data points are for objects likely affected by various bias factors including association with bright H II regions, SN remnants, strong X-ray emission, and the extragalactic objects in the north Galactic polar region with b > 45°. The data point for vZ 1128 at $\log|z| = 1$ is also plotted with an open circle.

FIG. 3- $\log[N(C\ IV)/N(O\ VI)]$, $\log[N(Si\ IV)/N(O\ VI)]$, and $\log[N(Si\ IV)/N(C\ IV)]$ are plotted versus $\log|z|$ and $\log d$. Filled circles are for the restricted sample. Open circles include objects likely to be affected contamination from associated Nebular regions, SNRs, wind blown bubbles, and for extragalactic directions(and vZ 1128) through the north Galactic polar region with b>45° which shows excess high ion column densities. Note the large ($\sim$ factor of 2) decrease in the dispersion of the ion ratios for the restricted sample when moving from lines of sight to stars in the Galaxy with $\log|z|$ and log d < 1 to the extragalactic lines of sight with $\log|z|$ and $\log d > 1$. To avoid confusion upper limits and data points with very large error bars are not plotted.

FIG. 4.- Galactic exponential scale heights for the WNM as traced by H I,  the WIM as traced by Al III,  and the ions that exist in the transition temperature plasma as traced by Si IV, C IV and  O VI.  The scale heights are plotted against energy (eV) required to produce and destroy the ion with the dashed horizontal lines displaying the energy ranges.   The results from this work are shown with the filled circles.   The value for N V displayed with the open circle from Savage et al. (1997) is based on measurements for a small sample of extragalactic lines of sight.

FIG. 5.- Values of N(C IV) versus log[N(Si IV)/N(C IV)] found in 17 different absorbing components along the lines of sight to 5 disk stars observed at ~3 km s$^{-1}$ resolution and high S/N are plotted with filled circles and filled squares.  The total interstellar path length is 20.5 kpc.  Components labeled NPS and SSS are associated with Galactic supershells.  The other 14 components show a wide range in the values of log[N(Si IV)/N(C IV)] with the stronger four components lying  in the range from log[N(Si IV)/N(C IV)] = -0.75 to -0.60.  Four weak components with C IV line widths implying gas with T <$5x10^4$ K are plotted with the filled squares. Photoionization in warm gas with T ~$10^4$ K is the likely origin of the ionization in these weak components.

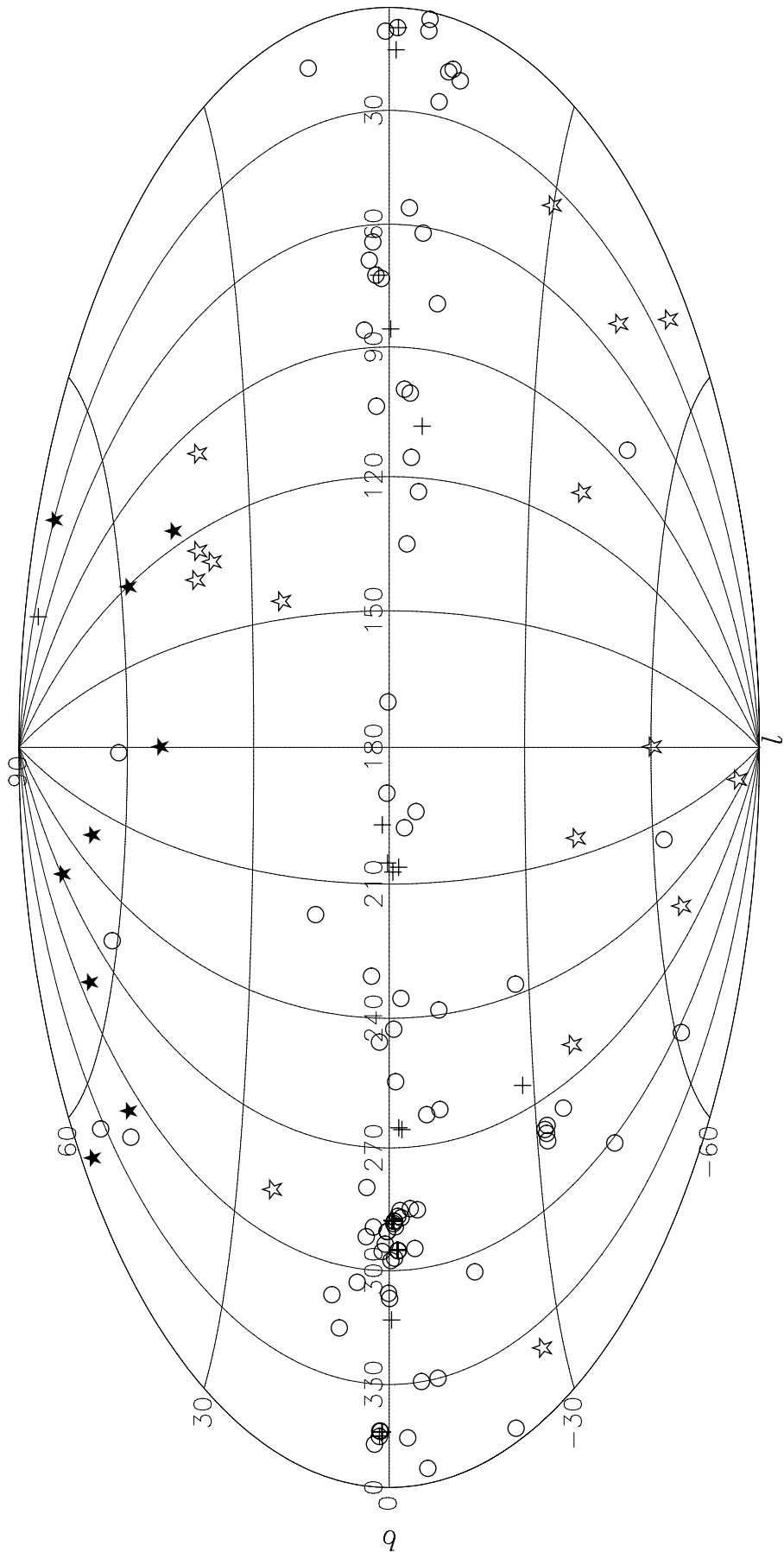

FIG. 1a

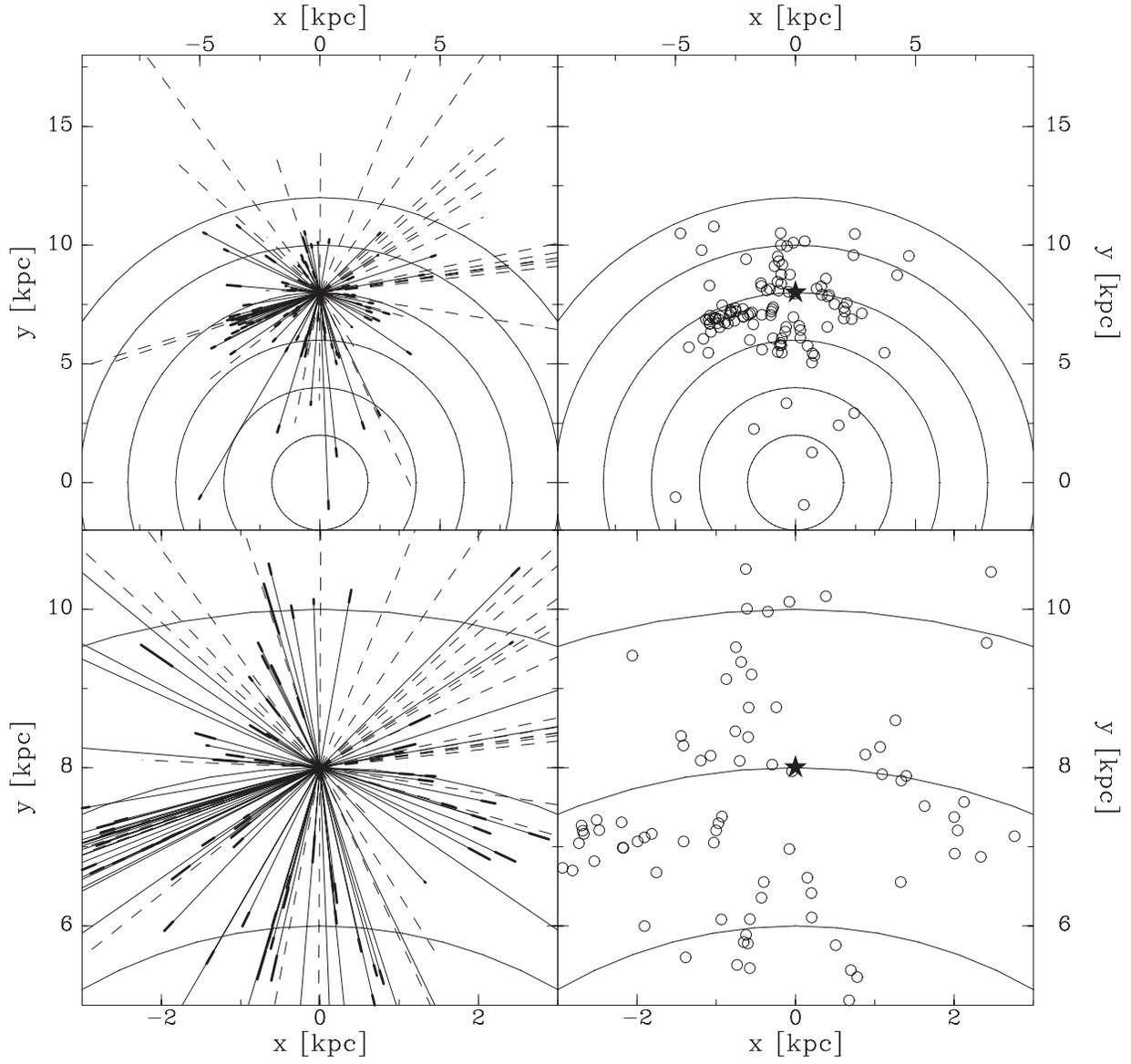

FIG. 1b

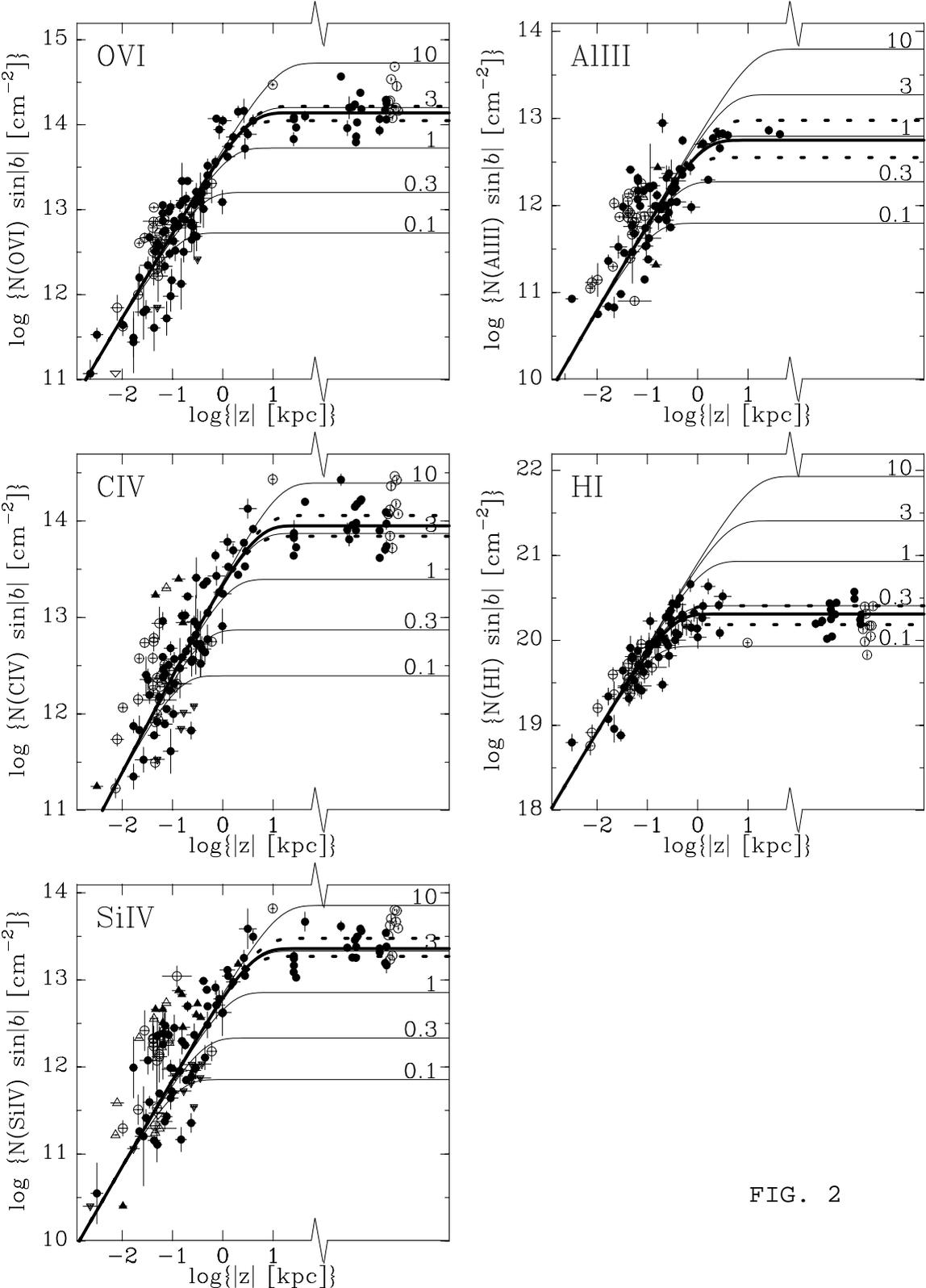

FIG. 2

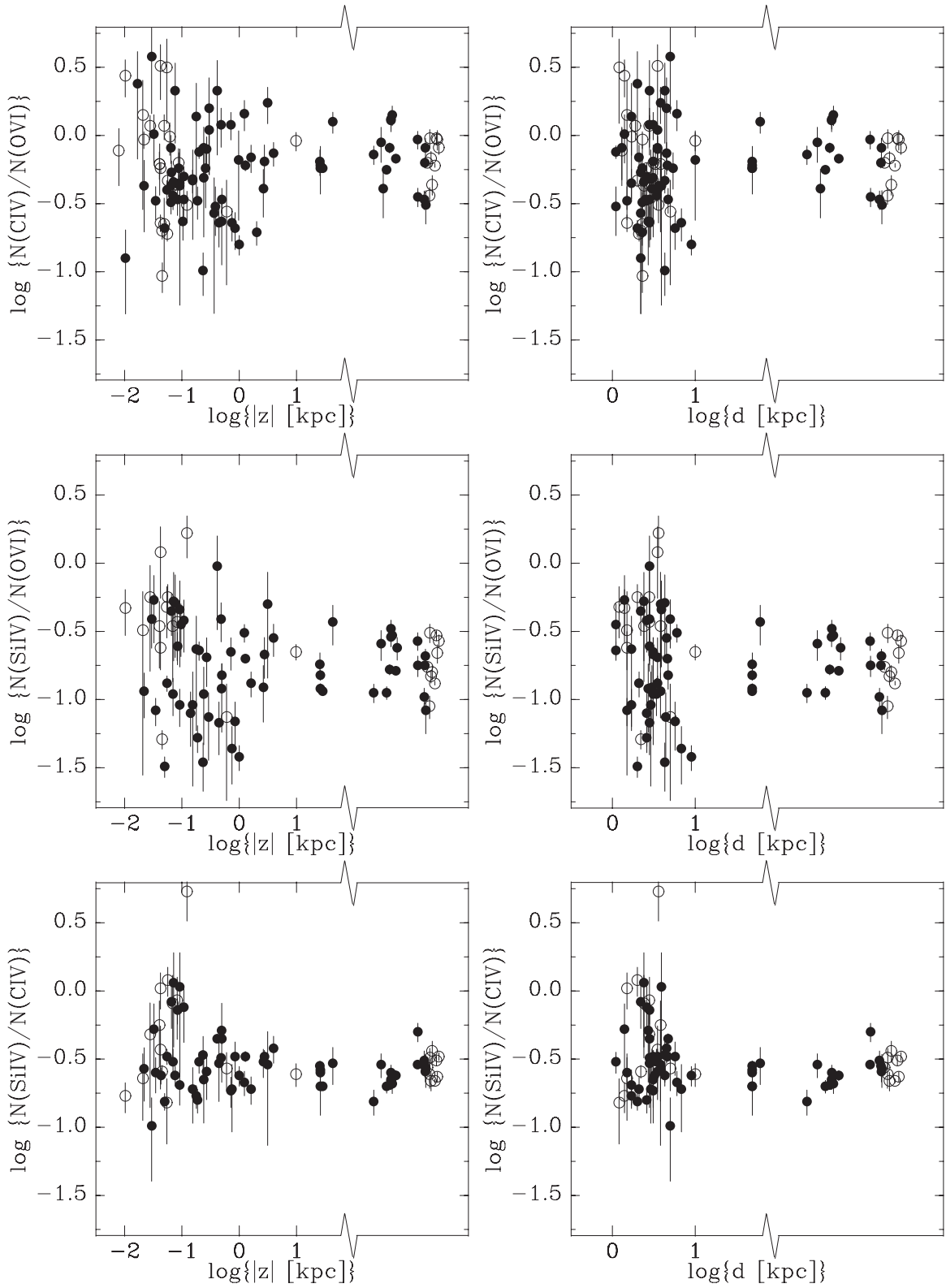

FIG. 3

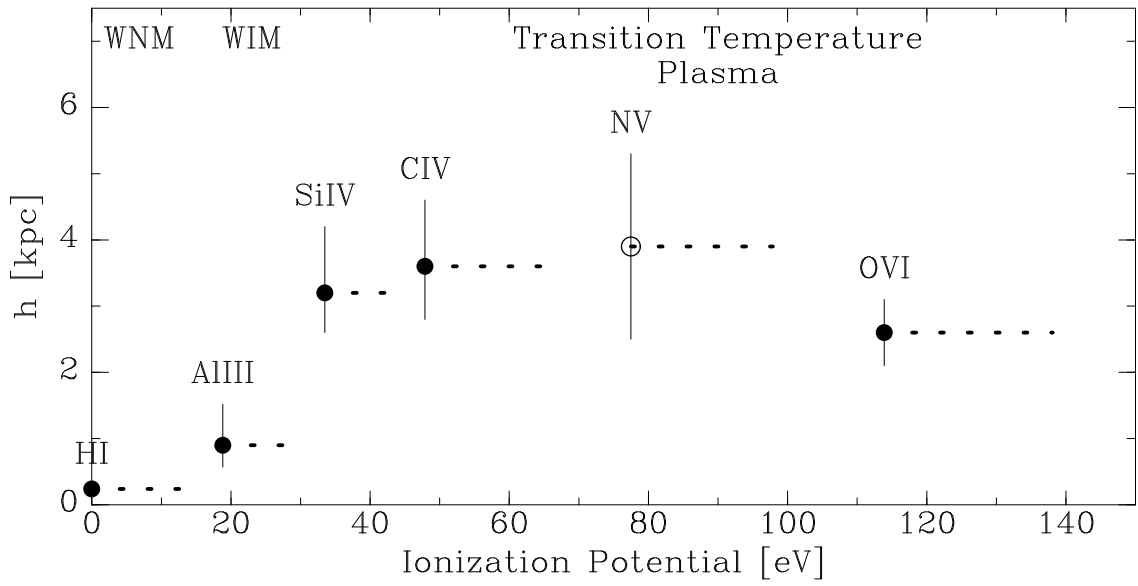

FIG. 4

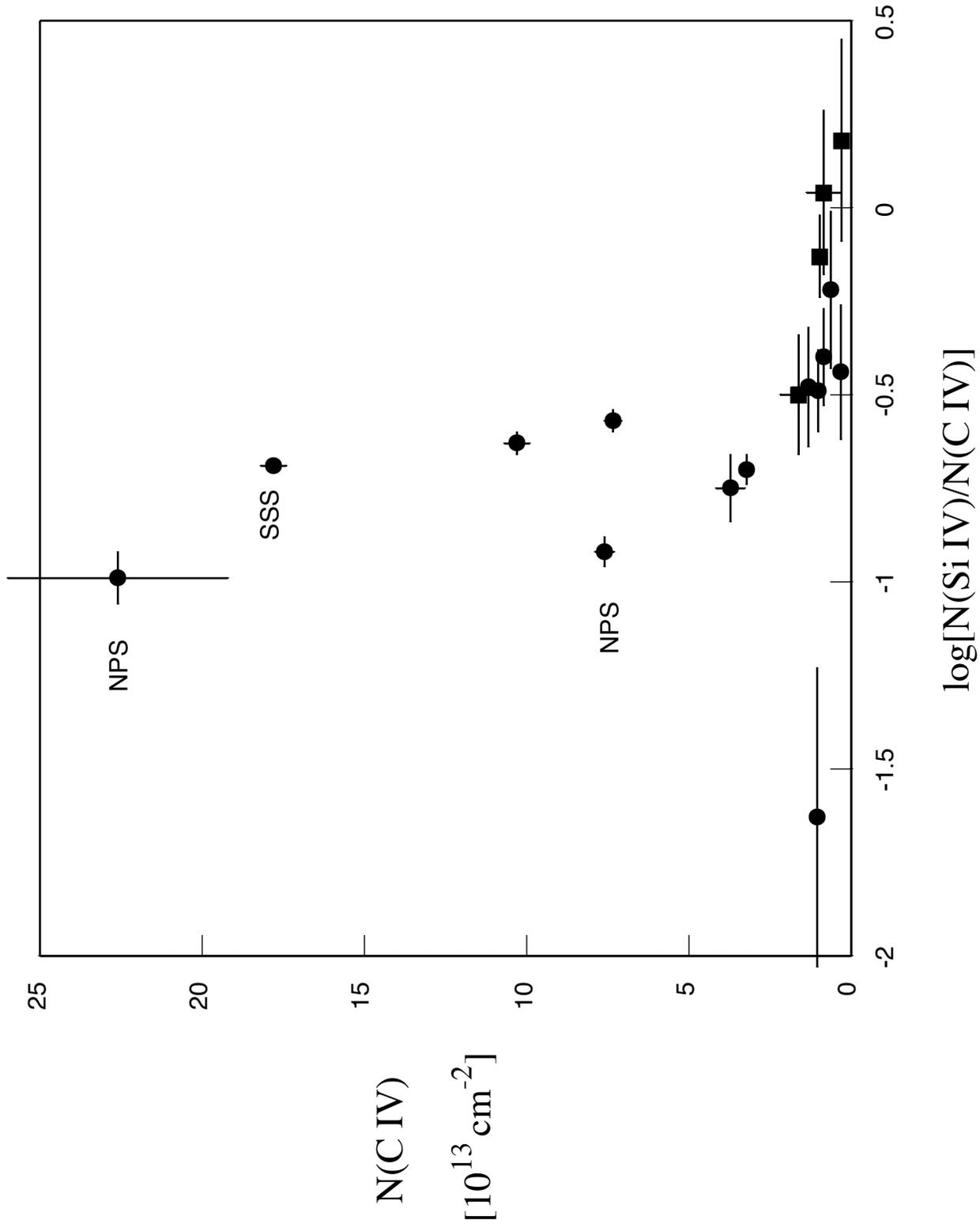

FIG. 5